\documentclass[onecollarge,smallextended]{svjour2} 
\usepackage{amsmath}
\usepackage{verbatim}
\usepackage{graphicx}
\usepackage{subfigure}
\usepackage{bm}
\begin{document}
\title{\Large{The theory of individual based discrete-time processes}}
\author{\large{Joseph~D. Challenger } \and
\large{Duccio Fanelli}   \and 
\large{Alan~J. McKane} 
}
\institute{JD Challenger \and D Fanelli \at 
Dipartimento di Fisica e Astronomia, Universit\`{a} degli Studi di Firenze and INFN, via Sansone 1, IT 50019 Sesto Fiorentino, Italy \\
\email{jchallenger@unifi.it} \\
\email{duccio.fanelli@unifi.it}
\and
AJ McKane \at 
Theoretical Physics Division, School of Physics and Astronomy, 
The University of Manchester, Manchester M13 9PL, UK\\
\email{alan.mckane@manchester.ac.uk}
}
\maketitle
\vspace{-0.8cm}

\begin{abstract}
A general theory is developed to study individual based models which are discrete in time. We begin by constructing a Markov chain model that converges to a one-dimensional map in the infinite population limit. Stochastic fluctuations are hence intrinsic to the system and can induce qualitative changes to the dynamics predicted from the deterministic map. From the Chapman-Kolmogorov equation for the discrete-time Markov process, we derive the analogues of the Fokker-Planck equation and the Langevin equation, which are routinely employed for continuous time processes. In particular, a stochastic difference equation is derived which accurately reproduces the results found from the Markov chain model. Stochastic corrections to the deterministic map can be quantified by linearizing the fluctuations around the attractor of the map. The proposed scheme is tested on stochastic models which have the logistic and Ricker maps as their deterministic limits. 
\end{abstract}
%\PACS{05.40.-a, 02.50.Ey, 05.45.-a.}
\section{Introduction}
\label{intro}
In many areas of science, numerical simulations have become the dominant methodology, often replacing more traditional methods of analysis, for instance, those based on phenomenological differential equations. One of the positive aspects of this change is that, when constructing models of specific systems, the focus has shifted from phenomenological models at the macroscale to individual based models (IBMs) at the microscale. This shift in emphasis is also important when carrying out analytical work: a consistent procedure consists of correctly formulating an IBM, and systematically deriving the more coarse-grained mesoscopic or macroscopic descriptions from these. Although this is commonly carried out in the physical sciences, it is far less common in the biological sciences. 

The careful setting-up of IBMs is especially important in the cases where the intrinsic noise, arising from the discrete nature of the individuals, gives rise to structures which are seen on the macroscale \cite{mckane_05,butler_09,biancalani_10}. The structure of these depend on the nature of the IBM and are absent in the (deterministic) macroscale equations. We stress here that by ``microscale'' we mean the level of description which treats the fundamental entities of the system (be they atoms and molecules in physical systems or individuals in biological systems) as the basic quantities from which the model is constructed. Such models will be intrinsically stochastic with the time evolution of relevant averaged quantities giving the macroscopic description.

In the case of continuous-time models, this procedure is routinely carried out \cite{black_12,goutsias_13}. The IBM is formulated in terms of reactions between the basic entities, which can be both simulated using a Gillespie-type algorithm \cite{gillespie_76,gillespie_92book} or mathematically described as a master equation \cite{kampen_07}. The IBM in this latter form is intractable in all cases of interest, but progress is possible by making the approximation that the discrete variables describing the system can be replaced by continuous ones --- yet keeping the stochastic element. This gives a mesoscopic description in the form of either a Fokker-Planck equation or a stochastic differential equation. The macroscopic model is defined by the equation for the mean of the probability distributions found from these equations \cite{gardiner_09,risken_89}.

As far as we are aware, the procedure just described has not been explicitly carried out for discrete-time processes, even though the phenomenological modeling of biological processes through nonlinear maps is standard (see e.g. Refs.~\cite{may_75,hassell_75,may_76,strogatz_94}). The purpose of this paper is to develop such a formalism, and to apply it to simple models to explore the basic structure of such systems. Preliminary results of our study have already been published in a short communication \cite{challenger_13}. Here, we will give a more complete treatment, concentrating on the derivation of the equations governing the stochastic dynamics. In effect we are asking how discrete-time IBMs in the limit $N \to \infty$, where $N$ is the number of basic entities in the system, lead to nonlinear maps. We would expect to begin from discrete-time, discrete state-space IBMs in the form of Markov chains, which would then be approximated for large $N$ to give mesoscopic stochastic
  difference equations, rather than stochastic differential equations. One advantage of investigating these systems, rather than those with continuous time, is that chaotic behavior exists with only one degree of freedom, rather than requiring at least three if time is continuous \cite{strogatz_94}. Therefore the interplay of stochasticity and chaos should be more amenable to investigation in models where time is discrete. 

In Section 2 of the paper we will develop the formalism required to analyze the transition from discrete time, discrete state-space stochastic processes to discrete-time, continuous state-space processes. There are analogues of the Fokker-Planck equation and stochastic differential equation (now a stochastic difference equation) in the mesoscopic description and we derive these and explore their structure. In Section 3, we show that a linearization of the stochastic difference equation can, in certain circumstances, allow analytical progress to be made. Section 4 contains some results from the theory developed in earlier sections, using the logistic map as a case study. We also show similar results for an alternative map with a single degree of freedom, demonstrating the generality of the formalism we have constructed. Additionally, Section 4 contains an investigation of the interplay between stochasticity and chaos. We finish with a summary and outline of further work.
\section{Formalism}
\label{formal}
Although the formalism we will construct in this section holds quite generally for discrete-time Markov processes, we will not present it in its fully generality, since the important aspects risk being obscured by extra detail. So, for instance, we will restrict ourselves to systems with one variable. The essential points are already evident in this case; the extension to more than one variable is straightforward and will be discussed elsewhere. We will also use the language of ecology, since in this field there is an extensive literature on the construction of discrete-time models describing the interaction between individuals. However it should be clear from our discussion that the ideas are more widely applicable. The IBM we use in a given situation will of course be defined by the application being considered, but to establish the formalism we need to choose a particular class of models to work with. Fortunately, ecological models involving only one variable frequently have the number of individuals of a single species as the stochastic variable, and competition for resources as the sole mechanism (apart from birth and death) which changes this variable. This specifies the class of IBM we will study; they will be competition models and will have the form introduced by us in a recent publication \cite{challenger_13}.
\subsection{Construction of the IBM}
\label{construction}
Suppose that there is a population of $m$ individuals of the same species residing in a given patch at time $t$. The number of individuals changes stochastically due to random births, deaths and competition for resources. Although this is frequently modeled as a continuous-time Markov process, there is a long tradition in ecology of assuming that the number of individuals is only defined at discrete time intervals (see e.g. Refs.~\cite{may_75,hassell_75,may_76,hassell_76,godfray_89,neubert_92}). These discrete time periods could correspond to one generation, an annual cycle, a daily cycle, or any other biologically relevant interval.

The individuals are assumed to be well-mixed, in other words, there is no spatial or other structure which picks out one or more of the individuals as being special. We denote the probability that an individual in the patch at time $t$ gives rise to an individual in the patch at time $t+1$ by $p$. This will be a function of $m$ and is where the biological modeling enters. The new population at time $t+1$ is then found by sampling the population at time $t$, with $Q_{nm}$ being the probability that there are $n$ individuals in the population at time $t+1$, given there were $m$ at time $t$.

Let us give a simple example in order to clarify this procedure. A simple choice for $p$ might be 
\begin{equation}
p = \lambda\,\left( \frac{m}{N} \right) \left( 1 - \frac{m}{N} \right),
\label{logistic_p}
\end{equation}
where $N$ is the carrying capacity of the patch (the maximum number of individuals that the patch can sustain) and $\lambda$ is a constant. This has the correct properties: (i) for $m \ll N$, the growth is exponential, with $\lambda$ being the difference between the birth rate and death rate, (ii) as $m$ increases, competition reduces population growth, until when $m=N$, the population collapses.
 
Since the population is well-mixed, a reasonable assumption is that the probability $Q_{nm}$ follows a binomial distribution:
\begin{equation}
Q_{nm} = {N \choose n} p^{n} \left( 1 - p \right)^{N-n},
\label{binomial}
\end{equation}
where $p$ is a function of $m/N$ given by Eq.~(\ref{logistic_p}). The choices (\ref{logistic_p}) and (\ref{binomial}) define the model. If the probability that the population has $n$ individuals at time $t+1$ is denoted by $P_{n,t+1}$, then 
\begin{equation}
P_{n,t+1} = \sum^{N}_{m=0} Q_{nm} P_{m,t}.
\label{Markov}
\end{equation}

At this stage, we should briefly mention the notation that we are using. Since both the time $t$ and the state label (e.g. $n$ or $m$) is a number, we will usually indicate them by a subscript, such as $P_{m,t}$, rather than as $P_m(t)$, which we would use if $t$ was continuous. However we will not slavishly adhere to this notation if it means a loss in clarity. For instance we will briefly discuss the Chapman-Kolmogorov equation which relates pdfs which depend on two times. There it is clearer to keep time as an argument of the function, even though it is discrete.

One calculation which is straightforward to perform is to find the $N \to \infty$ deterministic dynamics of the model. This is carried out by finding the equation for the time evolution of mean quantities:
\begin{eqnarray}
\langle n_{t+1} \rangle &\equiv& \sum_{n=0}^{N} n P_{n,t+1} \nonumber \\
&=& \sum_{n=0}^{N} n \sum^{N}_{m=0} Q_{nm} P_{m,t} \nonumber \\
&=& \sum_{m=0}^{N} \left\{ \sum^{N}_{n=0} n Q_{nm} \right\} P_{m,t} \nonumber \\
&=& \sum_{m=0}^{N} N p\left( \frac{m}{N} \right) P_{m,t},
\label{average_1}
\end{eqnarray}
since the average of the binomial distribution is $Np$. Dividing by $N$ and letting $N \to \infty$ we have 
\begin{eqnarray}
z_{t+1} \equiv \lim_{N \to \infty} \frac{\langle n_{t+1} \rangle}{N} &=&
\lim_{N \to \infty} \sum_{m=0}^{N} p\left( \frac{m}{N} \right) P_{m,t} \nonumber \\
&=& \lim_{N \to \infty} \left\langle p\left( \frac{n_t}{N} \right) \right\rangle 
\nonumber \\
&=& p\left( \left\langle \frac{n_t}{N} \right\rangle \right), 
\label{average_2}
\end{eqnarray}
the last line following from the fact that as $N \to \infty$ the dynamics are deterministic and the pdfs are delta-functions. Then, for instance, $\langle n^{\ell} \rangle = (\langle n \rangle )^{\ell}$. 
Therefore the deterministic dynamics is given by $z_{t+1}=p(z_t)$, or in the specific case (\ref{logistic_p}) which we are considering here,
\begin{equation}
z_{t+1} = \lambda z_t \left( 1 - z_t \right),
\label{logistic_map}
\end{equation}
which is the well known logistic map \cite{strogatz_94}.

It is interesting to ask how one should deduce a form for $p$, which in the example above is given by Eq.~(\ref{logistic_p}). In continuous time models, there is usually a mechanistic basis to the form of the interactions at the microscale, but in the case of discrete-time models this is likely to be an effective interaction reflecting the many and varied interactions that have occurred during the time interval. Thus the choice is phenomenological and is essentially of the same type as is made for deterministic discrete-time models. When there is only one variable the choice is quite restricted, but for many variables a greater range of mechanisms are possible.

\subsection{Derivation of the mesoscopic description}
\label{FPE}
In the mesoscopic description we will use the variable $z=n/N$, instead of $n$. One of the approximations that is made in going from the IBM to the mesoscopic formulation is that $z$ is assumed to be continuous, and so we expect the approximation to be good when $N$ is large. We would also expect that there should be an equation for $P_t(z)$ --- the analogue of $P_{n,t}$ --- which is similar to the Fokker-Planck equation for continuous time processes.

The most systematic way to derive this equation is to proceed as in the case of continuous time and to derive the Kramers-Moyal expansion \cite{gardiner_09}. The jump moments specific to the process defined by Eqs.~(\ref{logistic_p}) and (\ref{binomial}) are then substituted into the Kramers-Moyal expansion and truncated by neglecting $\mathcal{O}(N^{-2})$ terms. Although this procedure can be followed, we will see that for the processes of interest to us here there is a significant difference from the usual case.

We begin with the condition which defines a discrete-time continuous state-space Markov process, that is, the Chapman-Kolmogorov equation \cite{kampen_07} 
\begin{equation}
P(z,t+1|z_0,t_0) = \int dz' P(z,t+1|z',t) P(z',t|z_0,t_0).
\label{CK_eqn}
\end{equation}
The idea behind the derivation of the Kramers-Moyal expansion is that $P(z,t+1|z',t)=\langle \delta(z_{t+1} - z_t) \rangle_{z_{t}=z'}$ only involves one time step, and so can be calculated directly from Eqs.~(\ref{logistic_p}) and (\ref{binomial}). Then Eq.~(\ref{CK_eqn}) can give us the time-evolution of the system for all $t$. The derivation is found in textbooks \cite{gardiner_09}, however for convenience we also give it in appendix A. There it is shown that under quite mild assumptions, Eq.~(\ref{CK_eqn}) becomes
\begin{equation}
P(z,t+1|z_0,t_0) = \sum_{\ell=0}^{\infty} \frac{(-1)^{\ell}}{\ell!} 
\frac{\partial^{\ell} }{\partial z^{\ell}} \left[ M_{\ell}(z)
P(z,t|z_0,t_0) \right],
\label{KM_conventional}
\end{equation}
where $M_{\ell}(z)$ is the $\ell$th jump moment and is defined by
\begin{equation}
M_{\ell}(z) = \langle \left[z_{t+1} - z_{t}\right]^{\ell} \rangle_{z_{t}=z}.
\label{defn_M}
\end{equation}
Using the IBM to calculate $M_{\ell}(z)$ it is usually found that $M_{\ell}$ for $\ell > 2$ fall off faster than those for $\ell \leq 2$, and the series in Eq.~(\ref{KM_conventional}) can be truncated at $\ell = 2$, leading to the Fokker-Planck equation \cite{gardiner_09}.

In contrast to this situation, we find that if we use Eqs.~(\ref{logistic_p}) and (\ref{binomial}) to calculate $M_{\ell}$ then there are terms which cannot be neglected at order $\ell > 2$, in fact they cannot be neglected for any $\ell$. The reason for this is not hard to find: in continuous time processes the jumps $z(t) \to z(t+dt)$ are small; here the jumps $z_t \to z_{t+1}$ need not be. However, we would expect that the jumps $p_{t+1} \to z_t$ would be small, since this quantity is exactly zero on average, and so is equal to random variable with a variance proportional to $N^{-1}$. Therefore we write Eq.~(\ref{defn_M}) as
\begin{equation}
M_{\ell}(z) = \langle \left[z_{t+1} - p_t + 
\left\{ p_t - z_{t} \right\} \right]^{\ell} \rangle_{z_{t}=z},
\label{new_M}
\end{equation}
where it is understood that $p_{t}=p(z_t)$. We may now expand Eq.~(\ref{new_M}) to write $M_{\ell}(z)$ as
\begin{eqnarray}
M_{\ell}(z) &=& \sum^{\ell}_{r=0}\, {\ell \choose r} 
\left( p - z \right)^{\ell - r} \left\langle \left[ z_{t+1} - p_t \right]^r 
\right\rangle_{z_{t}=z} \nonumber \\
&=& \sum^{\ell}_{r=0}\, {\ell \choose r} 
\left( p - z \right)^{\ell - r} J_r(p),
\label{M_to_J}
\end{eqnarray}
where
\begin{equation}
J_r(p) \equiv \left\langle \left[ z_{t+1} - p_t \right]^r \right\rangle_{z_{t}=z},
\label{J_r}
\end{equation}
are the jump moments which are more natural in the context in which we are working.

The $J_{r}$ can now be evaluated using the properties of the IBM. As usual $J_0=1$, and $J_1$ is given by
\begin{eqnarray*}
\left\langle \left[ z_{t+1} - p_t \right] \right\rangle_{z_{t}=z} &=& 
N^{-1}\,\left\langle n_{t+1} \right\rangle_{n_{t}=m} - p \nonumber \\
&=& N^{-1}\,\sum_{n} n P(n,t+1|m,t) - p \nonumber \\
&=& N^{-1}\,\sum_{n} n Q_{nm} - p \nonumber \\
&=& N^{-1}\,N p - p = 0.
\end{eqnarray*}
A similar calculation gives $J_2(p)=N^{-1}p(1-p)$. In appendix B we show that $J_{r}=\mathcal{O}(N^{-2})$ for $r > 2$, and so only the jump-moments with $r \leq 2$ need be retained. Substituting these values for the $J_r$ into Eq.~(\ref{M_to_J}) gives
\begin{eqnarray}
& & M_{\ell}(z) = \left( p - z \right)^{\ell} + \frac{\ell(\ell - 1)}{2N}\,
\left( p - z \right)^{\ell - 2} p \left( 1 - p \right) + \mathcal{O}(N^{-2}).
\nonumber \\
\label{M_explicit}
\end{eqnarray}
We can now substitute Eq.~(\ref{M_explicit}) into the Kramers-Moyal expansion (\ref{KM_conventional}) to obtain
\begin{eqnarray}
& & P_{t+1}(z) = \sum_{\ell=0}^{\infty} \frac{(-1)^{\ell}}{\ell!} 
\frac{\partial^{\ell} }{\partial z^{\ell}} \left[ \left( p - z \right)^{\ell}
P_t(z) \right] \nonumber \\
& & + \frac{1}{2N}\,\sum^{\infty}_{\ell = 2}\,\frac{(-1)^{\ell}}{(\ell - 2)!}\,
\frac{\partial^{\ell}}{\partial z^{\ell}} \left[ (p-z)^{\ell - 2}\,p(1-p)\,
P_t(z)\right] + \mathcal{O}\left(\frac{1}{N^2}\right) \nonumber \\
& & = \sum_{\ell=0}^{\infty} \frac{(-1)^{\ell}}{\ell!} 
\frac{\partial^{\ell}}{\partial z^{\ell}} \left[ (p-z)^{\ell}\,P_t(z)\right]
\nonumber \\
& & + \frac{1}{2N}\,\sum^{\infty}_{\ell = 0}\,\frac{(-1)^{\ell}}{\ell !}\,
\frac{\partial^{\ell}}{\partial z^{\ell}}\,\frac{\partial^2}{\partial z^2}\,
\left[ (p-z)^{\ell}\,p(1-p)\,P_t(z)\right] + 
\mathcal{O}\left(\frac{1}{N^2}\right), \nonumber \\
\label{strange_KM_expansion}
\end{eqnarray}
where we have now not indicated the initial conditions on the pdfs explicitly and so have returned to the notation where time is indicated by a subscript on the pdfs.

Equation (\ref{strange_KM_expansion}) has a far more complicated form than the conventional Fokker-Planck equation, which only contains first and second order derivatives. Even the deterministic ($N \to \infty$) equation
\begin{equation}
P_{t+1}(z) = \sum_{\ell=0}^{\infty} \frac{(-1)^{\ell}}{\ell!} 
\frac{\partial^{\ell}}{\partial z^{\ell}} \left[ (p-z)^{\ell}\,P_t(z)\right],
\label{KM_expansion_deter}
\end{equation}
contains derivatives to all orders.

It is useful to focus on the less complicated equation (\ref{KM_expansion_deter}), rather than (\ref{strange_KM_expansion}), to explain how to obtain a simpler and more transparent form. A clue how to proceed is that to move from the jump moments $M_{\ell}$ to the $J_r$, one makes a constant shift of $(p-z)$. This suggests that one works in terms of the Fourier transform
\begin{eqnarray}
\tilde{P}_{t+1}(k) &=& \int^{\infty}_{-\infty} dz\,e^{ikz}\,P_{t+1}(z) \nonumber \\
&=& \sum_{\ell=0}^{\infty} \frac{1}{\ell!}\,\int^{\infty}_{-\infty} dz\,e^{ikz}\,
\frac{\partial^{\ell}}{\partial z^{\ell}} \left[ (p-z)^{\ell}\,P_t(z)\right],
\label{FT_step1}
\end{eqnarray}
where $P_t(z)$ be defined to be zero for $z<0$ and $z>1$. Integrating by parts and then carrying out the sum we have
\begin{eqnarray}
\tilde{P}_{t+1}(k) &=& 
\sum_{\ell=0}^{\infty} \frac{1}{\ell!}\,\int^{\infty}_{-\infty} dz\,e^{ikz}\,
\left( ik \right)^{\ell} \left(p - z \right)^{\ell} P_t(z) \nonumber \\
&=& \int^{\infty}_{-\infty} dz\,e^{ikz}\,e^{ik(p-z)}\,P_t(z)
= \int^{\infty}_{-\infty} dz\,e^{ikp}\,P_t(z) \nonumber \\
&=& \int^{\infty}_{-\infty} dp\,e^{ikp}\,\mathcal{P}_t(p),
\label{FT_step2}
\end{eqnarray}
where the pdf $\mathcal{P}$ is defined by $P_t(z)dz=\mathcal{P}_t(p)dp$. Taking the inverse Fourier transform of Eq.~(\ref{FT_step2}) yields
\begin{equation}
P_{t+1}(z)=\mathcal{P}_t(z).
\label{Liouville}
\end{equation}
It should be noted that taking the inverse Fourier transform results in the argument of $\mathcal{P}_t$ being $z$, not $p$. Another point worth making is that, although the transformation between the variables $z$ and $p$ may not be $1-1$, the change of variables leading to Eq.~(\ref{Liouville}) can be justified. This is discussed further in appendix A.

Equation (\ref{Liouville}) is a deterministic equation for the time evolution, and so $P_{t+1}(z)=\delta(z_{t+1}-z)$. However since $z_{t+1}=p_t$, it follows that $P_{t+1}(z)=\delta(p_t - z)= \mathcal{P}_t(z)$. More importantly, the same idea of taking the Fourier transform can be used to simplify the finite $N$ equation (\ref{strange_KM_expansion}), and so elaborate on the role of stochastic fluctuations, beyond the deterministic limit. Following the general strategy discussed in appendix A, one ends with:
\begin{equation}
P_{t+1}(z) = {\cal P}_t(z) + \frac{1}{2N}\,\frac{\partial^2 }{\partial z^2} 
\left[ z \left( 1 - z \right) \mathcal{P}_t(z) \right] +
\mathcal{O}\left(\frac{1}{N^2}\right). 
\label{second_order_form}
\end{equation}

This is the equivalent of the Fokker-Planck equation for discrete time processes. To draw a connection with well-established results let us consider
a Wright-Fisher process where $p$ is simply $z$ \cite{ewens_04}. Equation (\ref{second_order_form}) reduces to 
\begin{equation}
P_{t+1}(z) - P_t(z) = \frac{1}{2N}\,\frac{\partial^2 }{\partial z^2} 
\left[ z \left( 1 - z \right) P_t(z) \right] +
\mathcal{O}\left(\frac{1}{N^2}\right),
\label{Wright_Fisher_FPE}
\end{equation}
which should be compared with the Fokker-Planck equation (for which the $\mathcal{O}(N^{-2})$ terms are discarded) for the Moran model:
\begin{equation}
\frac{\partial P(z,t)}{\partial t} = 
\frac{1}{2N}\,\frac{\partial^2 }{\partial z^2} 
\left[ z \left( 1 - z \right) P(z,t) \right] +
\mathcal{O}\left(\frac{1}{N^2}\right).
\label{Moran_FPE}
\end{equation}
This is expected, since the Moran model is the overlapping generation version (which is frequently approximated by a continuous time process) of the non-overlapping (discrete time) Wright-Fisher model \cite{ewens_04}.

Equation (\ref{second_order_form}) can be further generalized to account for higher order contributions. More precisely, as shown in appendix A, Eq.~(\ref{KM_conventional}) is found to be equivalent to 
\begin{equation}
P_{t+1}(z) = \sum_{r=0}^{\infty}\,\frac{(-1)^r}{r!}\,
\frac{\partial^{r}}{\partial z^{r}}\,\left[ J_r(z) \mathcal{P}_t(z) \right].
\label{KM_new}
\end{equation}
The difference between this latter equation and Eq.(\ref{KM_conventional}) is, of course, that the $J_r$ are $\mathcal{O}(N^{-2})$ for $r>2$, so that the expansion can be terminated at $r=2$, whereas the $M_{\ell}$ do not have this property.

\subsection{The stochastic difference equation}
\label{SDE}
The mesoscopic level of description of a continuous time process can either be expressed as a Fokker-Planck equation or as the equivalent stochastic differential equation \cite{gardiner_09,risken_89}. We would therefore expect a stochastic \textit{difference} equation to exist which would be equivalent to Eq.(\ref{second_order_form}). It should take the form $z_{t+1}=p_t \,+ $ noise, where the correlator of the noise should have the form appearing in the diffusion-like term in Eq.~(\ref{second_order_form}). Therefore we write it as
\begin{equation}
z_{t+1} = p_t + \eta_t,
\label{stoch_diff_eqn}
\end{equation}
where $\eta_t$ is a Gaussian variable with zero mean and correlator
\begin{equation}
\langle \eta_t \eta_{t'} \rangle = \frac{p(1-p)}{N} \delta_{t\,t'}.
\label{correlator}
\end{equation}
It should be noted that here the delta-function is a Kronecker delta, not a Dirac delta, and that the $\eta$ are not functions but random numbers drawn from a Gaussian distribution. 

To check that Eq.~({\ref{stoch_diff_eqn}) is equivalent to Eq.~(\ref{second_order_form}) is straightforward. Clearly, $\langle z_{t+1} \rangle = \langle p_t \rangle$, since $\langle \eta_t \rangle = 0$. Also,
\begin{equation}
J_2(p) = \left\langle \left[ z_{t+1} - p_t \right]^2 \right\rangle_{z_{t}=z} = 
\left\langle \eta^2_t \right\rangle = \frac{p(1-p)}{N},
\label{J_2_SDE}
\end{equation}
using Eq.~(\ref{correlator}). In addition, since $\eta_t$ is a Gaussian variable, $J_r(p) = \langle [ z_{t+1} - p_t ]^r \rangle_{z_{t}=z} = \langle \eta^r_t \rangle$ is of order $N^{-2}$ for $r > 2$. Substituting these values for $J_r$ in to Eq.~(\ref{KM_new}), shows that the Kramers-Moyal expansion for the process defined by Eqs.~(\ref{stoch_diff_eqn}) and (\ref{correlator}) is exactly that given by Eq.~(\ref{second_order_form}).

Equations (\ref{second_order_form}) and (\ref{stoch_diff_eqn}) are the mesoscopic equations describing the IBM defined by (\ref{logistic_p}) and (\ref{binomial}). They will form the starting point of the analysis described in Sections 3 and 4. For large values of $N$, when the numerical solution of the Markov chain becomes impractical, they provide a means of obtaining accurate results. The stochastic difference equation (\ref{stoch_diff_eqn}) can also be linearized about a fixed point (or $n$-cycle) to analytically investigate the nature of the stochastic fluctuations about these attractors. This will be the subject of the next section.

\section{The linearized equation}
\label{linearise}
In the previous section we obtained a stochastic difference equation that provides a mesoscopic description of the stochastic dynamics. As this equation is non-linear, it is difficult to study analytically. However, if the corresponding macroscopic equation has a stable fixed point, we can make progress by linearizing the stochastic difference equation around this point.

Denoting such a stable fixed point by $z^*$, we substitute 
$z_t=z^* +\xi_t/\sqrt N$ into the stochastic difference equation (\ref{stoch_diff_eqn}) and keep only the leading term in $1/\sqrt{N}$, or equivalently in $\xi$. This gives the linear stochastic difference equation
\begin{equation}
\xi_{t+1}=J\xi_t+\rho_t, \,\, \textrm{where}\,\, \langle \rho_t \rho_{t'} \rangle =B\delta_{t\, t'},
\label{linear_eqn}
\end{equation}
and where we have introduced $\rho_t = \sqrt{N} \eta_t$. Here $J \equiv \partial p/\partial z|_{z=z^*}$ and $B \equiv p^*(1-p^*)$ (the noise correlator for $\rho_t$). We can write $B=z^*(1-z^*)$, since $p=z$ at the fixed point. We also note that since we will be interested in expanding about a stable fixed point, we will ask that $|J| < 1$. 
Once an initial condition is given, this linear equation can be solved, by writing
\begin{eqnarray}
\xi_{t}&=&J\xi_{t-1}+\rho_{t-1}=J[J\xi_{t-2}+\rho_{t-2}]+\rho_{t-1} \nonumber \\
&=&J^2[J\xi_{t-3}+\rho_{t-3}]+J\rho_{t-2}+\rho_{t-1} = \hdots \nonumber \\
&=& J^m\xi_{t-m}+\sum_{k=0}^{m-1}J^k\rho_{t-(k+1)}, \ \ m=1,2,\ldots.
\label{lin_map_iter}
\end{eqnarray}
We impose the initial condition on $\xi$ to be $\xi_0$ at $t=t_0$. Therefore, taking $m=t-t_0$ in Eq.~(\ref{lin_map_iter}), we find the general solution to Eq.~(\ref{linear_eqn}) to be
\begin{equation}
\xi_t=J^{(t-t_0)}\xi_0 + \sum_{k=0}^{t-(t_0 + 1)}J^k\rho_{t-(k+1)}.
\label{gen_lin_soln}
\end{equation}
From here expressions for the first and second moments of $\xi_t$ can be found, using the properties of the noise: 
\[
\langle \xi_t \rangle=J^{(t-t_0)}\xi_0,
\]
and
\begin{equation}
\langle \xi^2_t \rangle = J^{2(t-t_0)}\xi_0^2+B\sum_{k=0}^{t-(t_0 +1)}J^{2k}= J^{2(t-t_0)}\xi_0^2+B\frac{1-J^{2(t-t_0)}}{1-J^2}.
\end{equation}
Here we are interested in studying fluctuations when the system is in a stationary state, and so we will focus on the moments in the limit $t_0 \to -\infty$, that is, when information about the initial condition has been forgotten. In this case 
\begin{equation}
\langle \xi \rangle^{\textrm{st}}=0, \ \ \ \langle \xi^2 \rangle^{\textrm{st}} = \frac{B}{1-J^2}.
\label{first_two_moments}
\end{equation}

A similar procedure may also be used to linearize about an $n$-cycle \cite{strogatz_94}, where $n\geq 2$. We begin with the 2-cycle, linearizing around the two points $z_1$ and $z_2$ that comprise the stable 2-cycle in the deterministic map. One extra complication is that we will have to evaluate both the $J$ and the noise correlators at two different values of $z$ ($z_1$ and $z_2$), rather than one ($z^*$). We will denote the two $J$ values by $J_1$ and $J_2$ and the noises by $\rho^{(1)}$ and $\rho^{(2)}$. We now relate $\xi_{t+2}$ to $\xi_t$ by applying Eq.~(\ref{linear_eqn}) twice. Starting the system near $z_1$ we have that
\begin{equation}
\xi_{t+2}=J_1J_2\xi_t+J_2\rho^{(1)}_{t}+\rho^{(2)}_{t+1}, \ \ \langle \rho^{(a)}_t \rho^{(a)}_{t'} \rangle \equiv B_a\delta_{t\, t'},
\label{linear_eqn_2}
\end{equation}
with $a=1,2$. This equation may be written in the more compact form
\begin{equation}
\xi_{t+2}=J\xi_t+\sigma^{(1)}_{t}, \ \ \ \sigma^{(1)}_t \equiv J_2\rho^{(1)}_{t}+\rho^{(2)}_{t+1},
\label{compact_eqn_2}
\end{equation}
where $J \equiv J_1J_2$. This has exactly the same form as the difference equation (\ref{linear_eqn}), except for the fact that $t$ changes by 2 during every iteration. So proceeding in the same way as for the case of fluctuations about the fixed point, one finds that 
\begin{equation}
\xi_{t} = J^m\xi_{t-2m}+\sum_{k=0}^{m-1}J^k\sigma^{(1)}_{t-2(k+1)}, \ \ m=1,2,\ldots.
\label{lin_map_iter_2}
\end{equation}
Taking $2m=t-t_0$, and letting $t_0 \to -\infty$ yields
\begin{equation}
\xi_{t} = \sum_{k=0}^{\infty}J^k\sigma^{(1)}_{t-2(k+1)}.
\label{lin_map_soln_2}
\end{equation}
From this equation we quickly find that $\langle \xi \rangle^{\textrm{st}}_1=0$ and 
\begin{equation}
\langle \xi^2 \rangle^{\textrm{st}}_1=\frac{B_1J_2^2+B_2}{1-(J_1J_2)^2} =
\frac{J^2}{(1-J^2)}\,\left\{ \frac{B_1}{J^2_1} + \frac{B_{2}}{(J^2_1 J^2_2)} 
\right\}.
\label{eq:2cycle_var1}
\end{equation}
Had we started around the other point in the 2-cycle, $z_2$, we would have found
\begin{equation}
\xi_{t+2}=J\xi_t+\sigma^{(2)}_{t}, \ \ \ \sigma^{(2)}_t \equiv J_1\rho^{(2)}_{t}+\rho^{(1)}_{t+1},
\label{0ther_compact_eqn_2}
\end{equation}
leading to Eq.~(\ref{lin_map_soln_2}), but with $\sigma^{(1)}$ replaced by $\sigma^{(2)}$. This yields $\langle \xi \rangle^{\textrm{st}}_2=0$ and 
\begin{equation}
\langle \xi^2 \rangle^{\textrm{st}}_2=\frac{B_2J_1^2 + B_1}{1-(J_1J_2)^2} =
\frac{J^2}{(1-J^2)}\,\left\{ \frac{B_2}{J^2_2} + \frac{B_{1}}{(J^2_2 J^2_1)} 
\right\}.
\label{eq:2cycle_var2}
\end{equation}

The general results for an $n$-cycle can be derived in a similar way. Suppose that the $n$-cycle is comprised of the points $z_1,z_2,\ldots,z_n$, and that $J$
and $B$ evaluated at these points are denoted by $J_1,J_2,\dots,J_n$ and $B_1,B_2,\ldots,B_n$ respectively. We relate $\xi_{t+n}$ to $\xi_t$ by applying Eq.~(\ref{linear_eqn}) $n$ times. Starting the system near $z_a$, $a=1,2,\ldots,n$, we have that
\begin{equation}
\xi_{t+n}=J\xi_t+\sigma^{(a)}_{t}, 
\label{compact_eqn_n}
\end{equation}
where $J=J_1J_2\ldots J_n$ and where
\begin{eqnarray}
\sigma^{(a)}_{t} &=& \rho^{(a-1)}_{t+n-1} + J_{a-1}\rho^{(a-2)}_{t+n-2} + J_{a-1}J_{a-2}\rho^{(a-3)}_{t+n-3} \nonumber \\
&+& \ldots + J_{a-1}J_{a-2}\ldots J_{a-(n-1)}\rho^{(a-n)}_{t}.
\label{new_iter_noise}
\end{eqnarray}
This means that for an $n$-cycle, $\sigma^{(a)}_{t}$ will contain $n$ terms.
Here labels are to be understood mod$n$, for instance, $a-(n-1)$ is equal to $a+1$. This again has exactly the same form as Eq.~(\ref{linear_eqn}), except for the fact that $t$ changes by $n$ during every iteration. So proceeding as before one finds that in the stationary state
\begin{equation}
\xi_{t} = \sum_{k=0}^{\infty}J^k\sigma^{(a)}_{t-n(k+1)},
\label{lin_map_soln_n}
\end{equation}
and so $\langle \xi \rangle^{\textrm{st}}_a=0$ and 
\begin{eqnarray}
\langle \xi^2 \rangle^{\textrm{st}}_a &=& \frac{J^2}{(1-J^2)}\,\left\{ \frac{B_a}{J^2_a} + \frac{B_{a+1}}{(J^2_a J^2_{a+1})} \right. \nonumber \\
&+& \left. \frac{B_{a+2}}{(J^2_a J^2_{a+1} J^2_{a+2})} + \ldots +
\frac{B_{a+n-1}}{(J^2_a \ldots J^2_{a+n-1})} \right\},
\label{eq:ncycle_vara}
\end{eqnarray}
where $a=1,2,\dots,n$ and where all labels are again mod$n$. For instance, when 
$n=3$ the variances about the points $z_1, z_2$ and $z_3$ of the 3-cycle are
\begin{eqnarray}
\langle \xi^2 \rangle^{\textrm{st}}_1 &=& \frac{J^2}{(1-J^2)}\,\left\{ \frac{B_1}{J^2_1} + \frac{B_{2}}{(J^2_1 J^2_{2})} + \frac{B_{3}}{(J^2_1 J^2_{2} J^2_{3})} \right\}, \nonumber \\
\langle \xi^2 \rangle^{\textrm{st}}_2 &=& \frac{J^2}{(1-J^2)}\,\left\{ \frac{B_2}{J^2_2} + \frac{B_{3}}{(J^2_2 J^2_{3})} + \frac{B_{1}}{(J^2_2 J^2_{3} J^2_{1})} \right\}, \nonumber \\
\langle \xi^2 \rangle^{\textrm{st}}_3 &=& \frac{J^2}{(1-J^2)}\,\left\{ \frac{B_3}{J^2_3} + \frac{B_{1}}{(J^2_3 J^2_{1})} + \frac{B_{2}}{(J^2_3 J^2_{1} J^2_{2})} \right\}.
\label{eq:3cycle_vars}
\end{eqnarray}
It should be noted, however, that as $n$ increases, the procedure is valid only for larger and larger $N$. The reason for this is that for an $n$-cycle, there will be $n$ Gaussian probability distributions, which will begin to overlap with each other as $n$ increases at fixed $N$. This can be avoided by increasing $N$ and so reducing the width of the distributions.

In the next section we will evaluate the theoretical results developed in the first part of the paper. For the most part, we will use the logistic map as a case study. 

\section{The logistic map}
To illustrate the ideas developed above, we examine the logistic map given by Eq.~\eqref{logistic_map}. We recall that in Eq.~\eqref{logistic_map}, $z_t$ represents the ratio of existing population to the maximum carrying capacity at discrete time $t$. For $0\leq \lambda \leq 4$, $z_t$ remains in the unit interval for all $t$. As $\lambda$ is changed, the map displays a variety of behavior. For $1< \lambda <3$ the map has a stable non-zero fixed point, then a cascade of period doubling bifurcations take place, up to the onset of chaos at $\lambda=3.56995\hdots$. Due to its nonlinearity, the map does not have a closed-form solution, except for particular values of $\lambda$, such as $\lambda=4$ \cite{strogatz_94}. As a result of its relative simplicity, the logistic map has been widely employed in the past as a test model to explore the concept of chaos. 

Working in this context, we will simulate the previously discussed discrete-time Markov chain model that converges to the logistic map in the deterministic limit. We will reconstruct the distribution function of fluctuations for different choices of the control parameter $\lambda$ and system size $N$. In doing so, we will demonstrate the accuracy of the stochastic difference equation~\eqref{stoch_diff_eqn} and examine the predictive ability of the linearized difference equation.
\subsection{The distribution of fluctuations}
In this section we demonstrate the validity of the stochastic difference equation as a mesoscopic description of the Markov chain dynamics. We do this by a direct comparison between the stationary distribution of the Markov chain \cite{reichl_80} and simulation results obtained by repeated iteration of the difference equation. Figures~\ref{fig:sde_mc_1} and \ref{fig:sde_mc_2} show the distribution of fluctuations for cases where the corresponding deterministic equation has a fixed point. Therefore, the fluctuations are located around the fixed point. In the case of the first figure, the fluctuations are well approximated by a Gaussian distribution, whereas the distribution of fluctuations in the second figure are noticeably skewed. In both cases, the stochastic difference equation captures the distribution extremely well. The value of $\lambda$ in Figure~\ref{fig:sde_mc_3} corresponds to a chaotic map in the deterministic equation. In contrast with the previous examples, the 
 probability distribution of the Markov chain is very broad, but is equally well-captured by the stochastic difference equation.
\begin{figure}
\begin{center}
\includegraphics[scale=0.85]{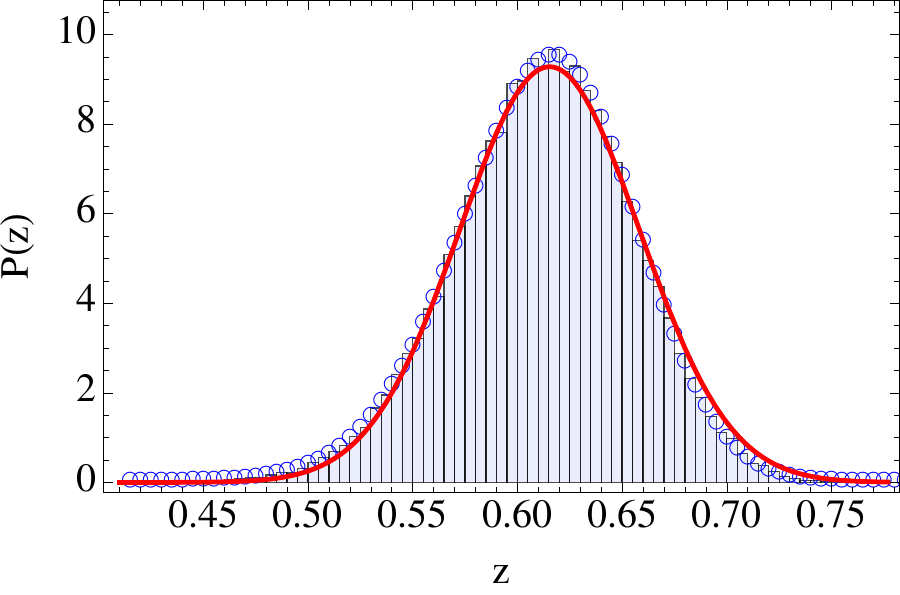}
\end{center}
\caption{Comparison between the stationary distribution of the Markov chain (blue circles) and simulations of the stochastic difference equation (bars) for $N=200$ and $\lambda=2.6$. The theoretical prediction from the linearized difference equation is also shown (red solid line). }
\label{fig:sde_mc_1}
\end{figure}
\begin{figure}
\begin{center}
\includegraphics[scale=0.85]{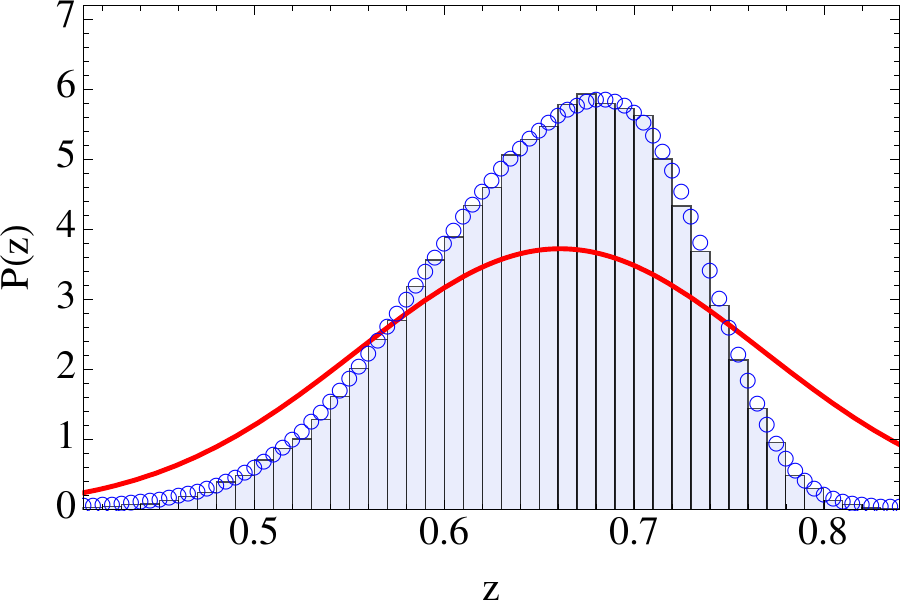}
\end{center}
\caption{Comparison between the stationary distribution of the Markov chain (blue circles) and simulations of the stochastic difference equation (bars) for $N=200$, $\lambda=2.95$. The theoretical prediction from the linearized difference equation is also shown (red solid line). }
\label{fig:sde_mc_2}
\end{figure}
\begin{figure}
\begin{center}
\includegraphics[scale=0.85]{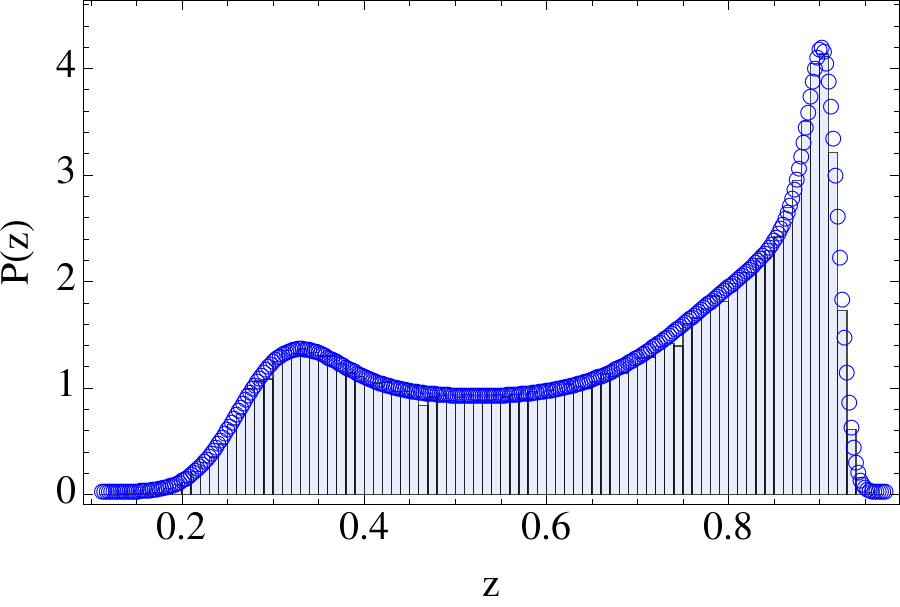}
\end{center}
\caption{Comparison between the stationary distribution of the Markov chain (blue circles) and simulations of the stochastic difference equation (bars) for $N=400$, $\lambda=3.65$. }
\label{fig:sde_mc_3}
\end{figure}
\subsection{Results from the linearized difference equation}
In Section~3, we obtained analytical expressions for the fluctuations by linearizing the stochastic equation around, for instance, a fixed point. Here we will use those equations to test the accuracy of this scheme using the logistic map.

For the case of the logistic map, given in Eq.~\eqref{logistic_map}, we can write the fixed point in terms of $\lambda$ as $z^*=1-1/\lambda$. The Jacobian evaluated at the fixed point is $J=\lambda (1-2z^*)=2-\lambda$. For the fixed point to be stable we require $|J|< 1$, or $1< \lambda <3$ \cite{strogatz_94}. So the variance in the stationary state, $\langle \xi^2 \rangle^{\textrm{st}}$ given in Eq.~(\ref{first_two_moments}), can be written as
\begin{equation}
\langle \xi^2 \rangle^{\textrm{st}}=\frac{B}{1-J^2}=\left(\frac{1}{\lambda}-\frac{1}{\lambda^2}\right)\frac{1}{1-(2-\lambda)^2}.
\end{equation}
The Gaussian distribution with this variance, and with a mean value of $z^*$, is shown for two parameter choices in Figures~\ref{fig:sde_mc_1} and \ref{fig:sde_mc_2}. In the former, it provides an excellent approximation to the stationary distribution of the Markov chain. However, in Figure~\ref{fig:sde_mc_2} it fails to capture the skewness of the stationary distribution. In general, the distribution can be expected to be skewed for small values of $N$, or near a bifurcation point. 

We also applied the linear theory to the case of the 2-cycle, using the expressions obtained in Eqs.~\eqref{eq:2cycle_var1} and \eqref{eq:2cycle_var2}. These results are compared to the stationary distribution of the Markov chain in Figure~\ref{fig:2cycle}, for $N=1400$ and $\lambda=3.2$ and show close agreement.
\begin{figure}
\begin{center}
\includegraphics[scale=0.9]{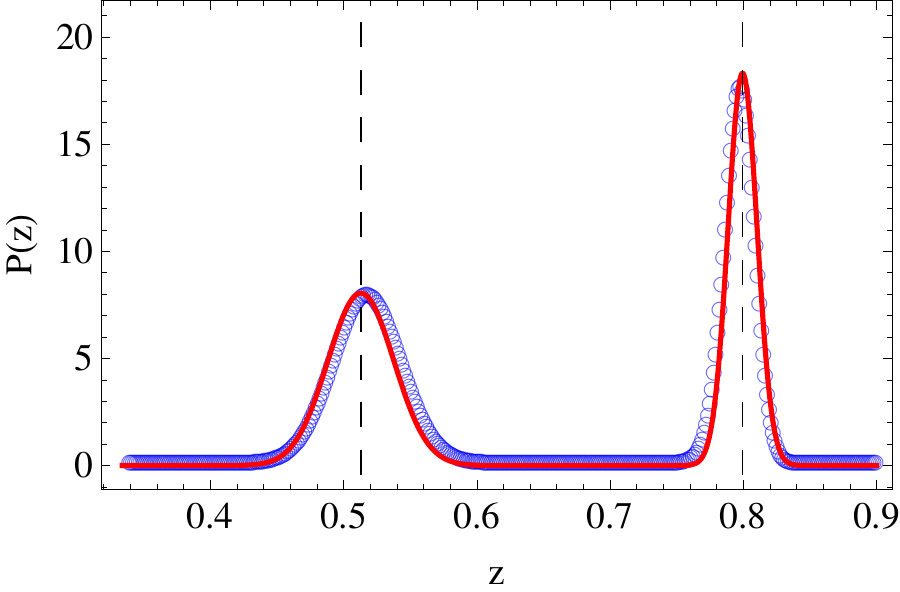}
\end{center}
\caption{Comparison between the stationary distribution of the Markov chain (blue circles) with the theoretical prediction obtained by linearizing around the 2-cycle (red solid line). The parameter values used were $N=1400$ and $\lambda=3.2$. The location of the 2-cycle in the deterministic equation is shown by the vertical dashed lines. }
\label{fig:2cycle}
\end{figure}

Finally, we show some results for the linearization around a 3-cycle, in Figure~\ref{fig:linearise_3_cycle}. Here we use a larger system-size ($N=250000$) since the 3-cycle is masked by noise, at low or intermediate $N$. Therefore, we compare the linearized solution as characterized by Eq.~(\ref{eq:3cycle_vars}), with simulation results from the full stochastic difference equation, rather than results from the Markov chain.
\begin{figure}
\begin{center}
\subfigure{\label{fig:3c1}\includegraphics[scale=0.75]{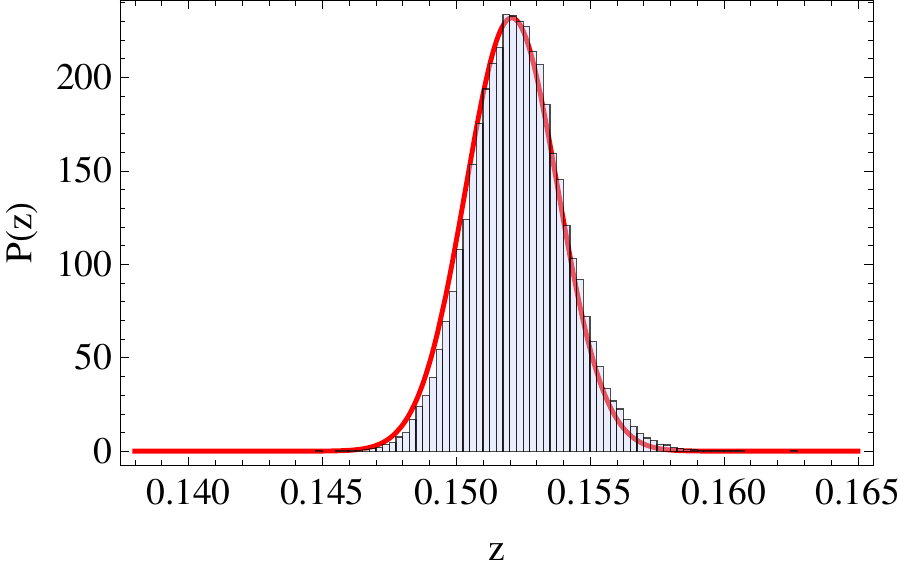}}
\subfigure{\label{fig:3c2}\includegraphics[scale=0.73]{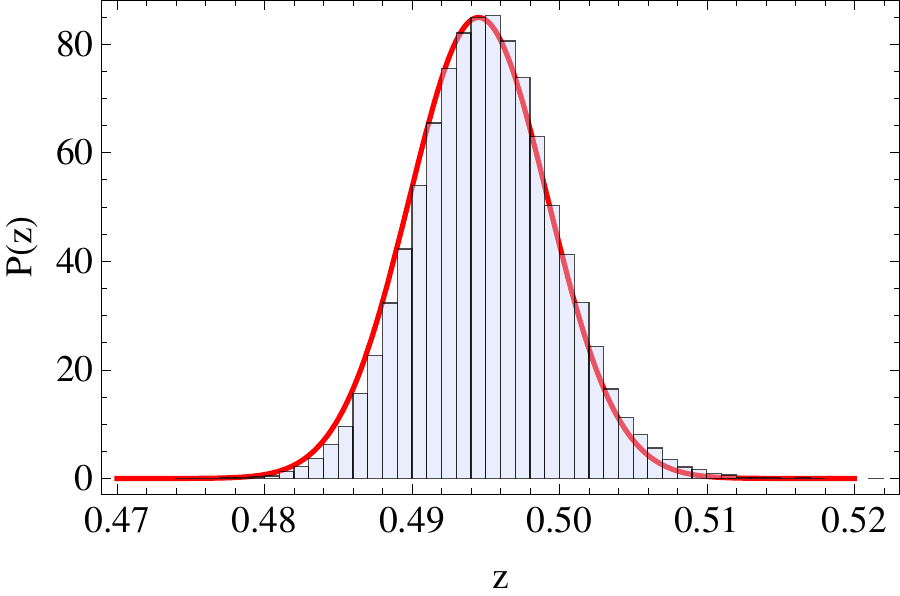}}\\
\end{center}
\begin{center}
\subfigure{\label{fig:3c3}\includegraphics[scale=0.74]{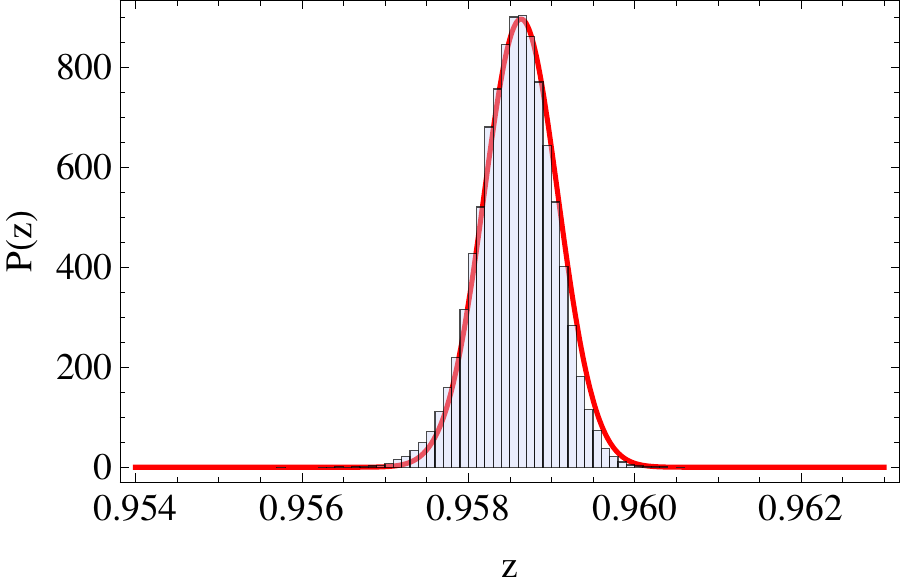}}
\end{center}
\caption{
Comparison between simulation results from the stochastic difference equation (bars) with the linearization around the 3-cycle (red line) for $\lambda=3.835$ and $N=250000$. For this choice of $\lambda$, the deterministic map has a 3-cycle at the points $z_1\simeq0.152$, $z_2\simeq0.495$ and $z_3\simeq0.959$.}
\label{fig:linearise_3_cycle}
\end{figure}
\subsection{Effect of noise on the bifurcation diagram}
When examining the role of fluctuations on the logistic map, it can be helpful to plot the bifurcation diagram, showing how the dynamics change as $\lambda$ is varied. We do this by simulating the stochastic difference equation, for a choice of $N$. As $N$ controls the strength of fluctuations, it will determine how much structure is preserved, compared with the bifurcation diagram of the deterministic system. Figure~\ref{fig:bif_sde} shows the bifurcation diagrams obtained for $N=8000$ (left) and $N=90000$ (right). For $N=8000$, structure beyond the 2-cycle is lost, including the deterministic windows beyond the onset of chaos. When $N=90000$, more structure is visible, including a 4-cycle, and a 3-cycle, found beyond the onset of chaos. To emphasize the contrast between these diagrams, we plot the distribution of points observed from simulations of the stochastic difference equation for a choice of $\lambda$ where the 3-cycle is stable in the deterministic equation. These distributions are shown in Figure~\ref{fig:prob_dists}, for $N=8000$ (left) and $N=90000$ (right). The transition from a narrow distribution to a broad one, as $N$ is decreases, demonstrates a clear change in the system's dynamics. The broad distribution is reminiscent of that found for a chaotic, deterministic map \cite{ott_93}. The nature of the dynamics represented by this distribution will be investigated in the next section. 
\begin{figure}
\begin{center}
\subfigure{\label{fig:bif_sde_N8000}\includegraphics[scale=0.8]{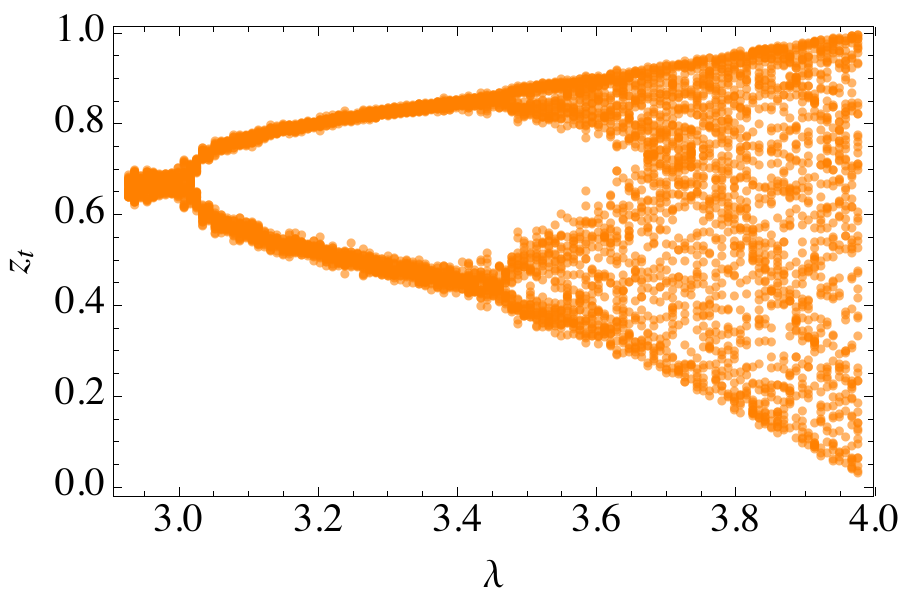}}
\subfigure{\label{fig:bif_sde_N90000}\includegraphics[scale=0.8]{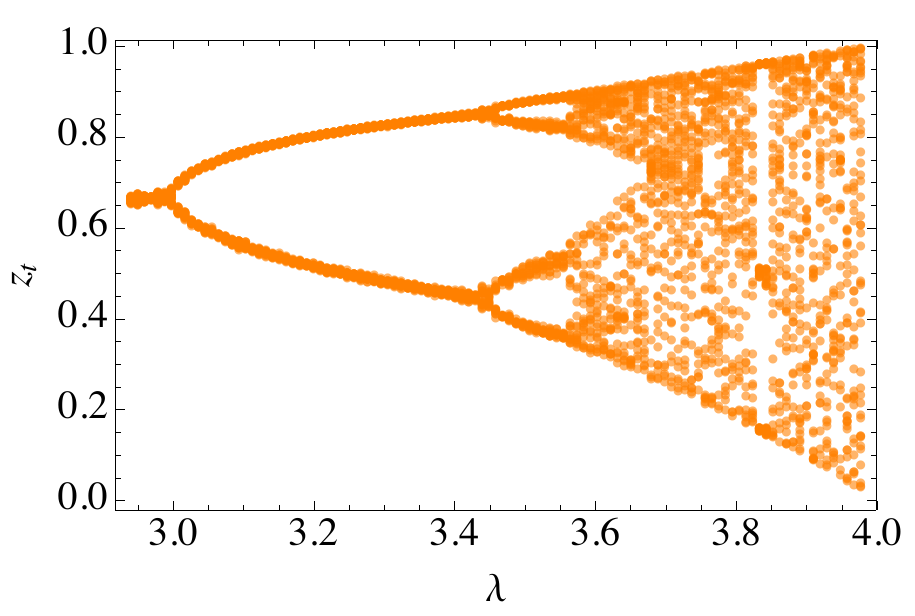}}
\end{center}
\caption{Bifurcation diagrams for the stochastic difference equation, obtained by simulating the stochastic difference equation with $N=8000$ (left) and $N=90000$ (right). For $N=8000$, structure beyond the 2-cycle regime is not preserved, due to the strength of the fluctuations. When $N=90000$, more structure is visible.}
\label{fig:bif_sde}
\end{figure}
\begin{figure}
\begin{center}
\subfigure{\label{fig:prob_dist_8000}\includegraphics[scale=0.8]{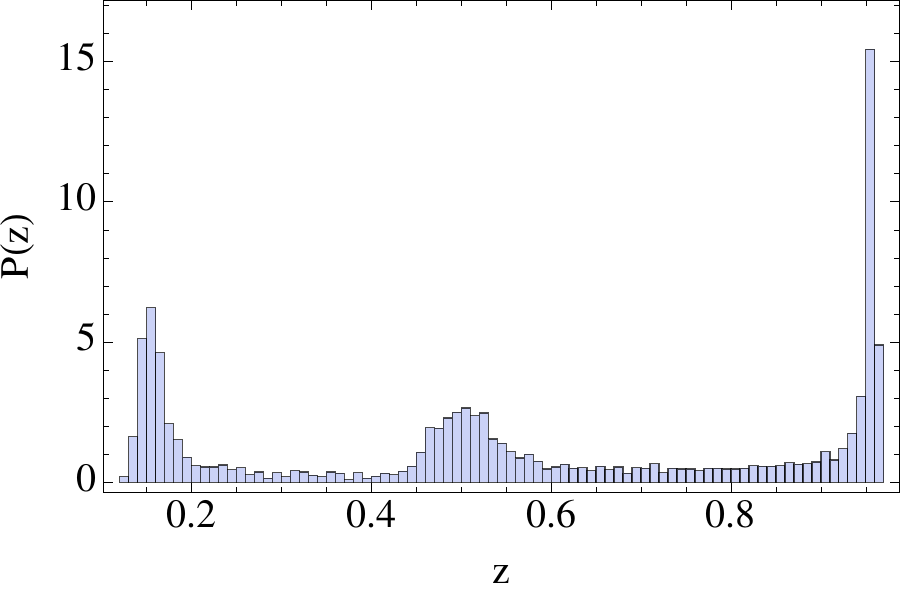}}\,\,\,\,
\subfigure{\label{fig:prob_dist_90000}\includegraphics[scale=0.8]{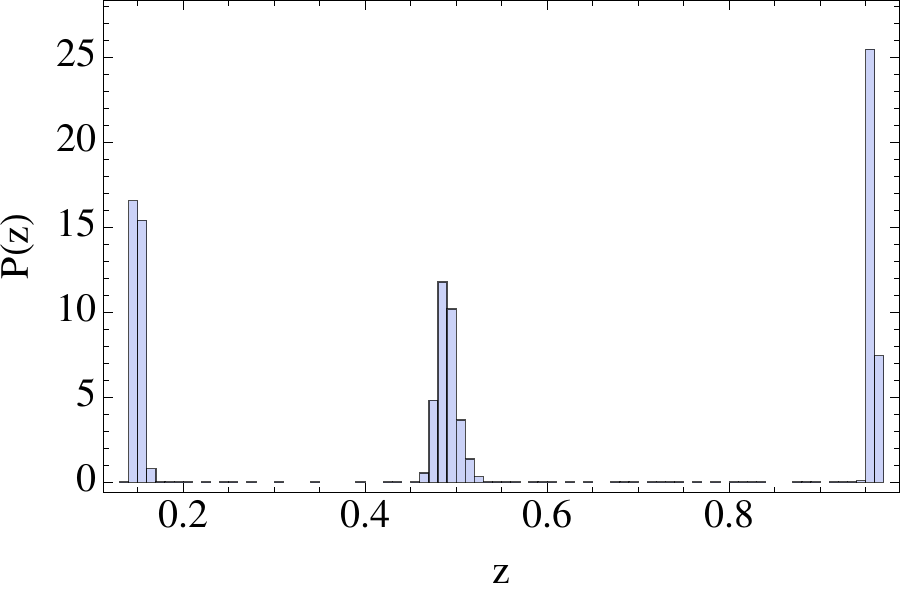}}
\end{center}
\caption{Probability distributions obtained by simulating the stochastic difference equation for $\lambda=3.84$. For this choice of $\lambda$ the deterministic map has a stable 3-cycle. The 3-cycle is not preserved for $N=8000$ (left), due to the strong noise, but it is visible for $N=90000$ (right).}
\label{fig:prob_dists}
\end{figure}
\subsection{Noise-induced chaos}
The ability of stochasticity to induce chaotic behavior has been much studied in both discrete and continuous time processes, but only for cases where the stochasticity originates from external perturbations to the system under consideration. A wide range of methods have been proposed to detect chaos in the presence of noise. Generally speaking, the methods used either involve examination of time series generated by the system, see e.g. Refs. \cite{gao_99,zunino_12,cencini_13}, or employ knowledge of the deterministic description of the process, see e.g. Refs. \cite{mayer_81,crutchfield_82}. Here we will investigate this effect for our intrinsic noise process, by estimation of the Lyapunov exponent, using one of each type of the two methods mentioned above. To measure the Lyapunov exponent, $\Lambda$, in the stochastic setting, we will first follow Refs.~\cite{mayer_81,crutchfield_82}, where a long iteration of the stochastic time series is fed into the formula for the Lyapunov exponent for a one-dimensional map, namely
\begin{equation}
\Lambda=\lim_{M\to\infty}\frac{1}{M}\sum_{i=1}^{M}\textrm{ln}|f'(z_i)|.
\label{eq:lyap}
\end{equation}
This method is very straightforward, and allows one to view how $\Lambda$ changes with $N$.
In Figure~\ref{fig:lyap_3cycle} we show the value of $\Lambda$ as a function of $N$ for $\lambda=3.8286$, which is chosen so that the corresponding deterministic map is non-chaotic (it has a stable 3-cycle), but is chaotic for nearby values of $\lambda$. In this figure, a noise-induced transition to chaos is predicted at around $N=2.4\times10^6$. Using this approach, the value of the Lyapunov exponent found for the corresponding deterministic map is recovered for $N$ sufficiently large.
\begin{figure}
\begin{center}
\includegraphics[scale=0.9]{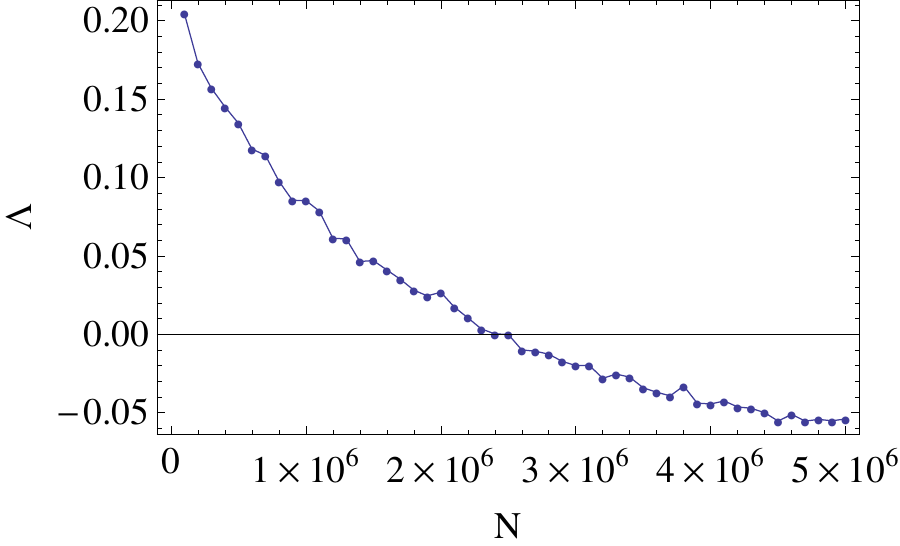}
\end{center}
\caption{Lyapunov exponent as a function of $N$ for $\lambda=3.8286$, calculated using Eq.~\eqref{eq:lyap}. For this value of $\lambda$ the deterministic map has a stable 3-cycle. }
\label{fig:lyap_3cycle}
\end{figure}

A key question to consider when discussing noise-induced chaos is how the analysis of the time series compares with that of a `clean', chaotic time series. This, of course, is particularly relevant when one analyses experimental data, where a description of the underlying dynamics is not known. Here we will analyze the simulation results of the stochastic difference equation using the time delay embedding technique \cite{packard_80}. This procedure is designed to reconstruct a system's attractor from its observed time series, a result formalized as Takens' theorem, which states that the attractor can be reconstructed if the embedding dimension is large enough \cite{takens_81}. Denoting a long time series from the stochastic difference equation by $x_i$, we define vectors $X_i$, such that $X_i=(x_i,x_{i+l},x_{i+2l},\hdots,x_{i+(m-1)l})$. Parameters $l$, the delay time, and $m$, the embedding dimension, must be suitably chosen \cite{ott_93,gao_94a}. In the work presented here, 
 we used $m=4$ and $l=1$. We follow the work of Gao et. al., who examined noise-induced chaos for an additive noise process \cite{gao_99}, by studying the time evolution of the system in the $m$-dimensional embedding space. They considered the form of time-dependent exponent curves, $L(k)$\footnote{Here we use $L(k)$ rather than $\Lambda(k)$, which was used by the authors in Ref.~\cite{gao_99}. This is to avoid confusion with $\Lambda$ used in Eq.~\eqref{eq:lyap}},
\begin{equation}
L(k)=\left\langle \textrm{ln} \left( \frac{||X_{i+k}-X_{j+k} ||}{||X_i-X_j||} \right) \right\rangle,
\end{equation}
where the averaging is performed over all pairs $(X_i,X_j)$, for which $||X_i-X_j||$ is found to lie within a prescribed small shell, denoted by $(r,r+\delta r)$. Calculating $L(k)$ over a range of shells, allows us to examine the system's behavior over a range of spatial scales. For deterministic chaos, the $L(k)$ curves will increase linearly before flattening. The curves from different scales collapse together as an envelope during the linear growth stage. The formation of this envelope has been used as a direct test for deterministic chaos, with its slope defining the value of the Lyapunov exponent \cite{gao_94a,gao_94b}. In Ref.~\cite{gao_99}, the authors considered the effect of additive Gaussian noise on the logistic map for different choices of $\lambda$ and noise strength. They found that, in certain cases, a noise-induced transition to chaos occurred. The question is, do we observe similar behavior for our intrinsic noise process?

Figure~\ref{fig:n_i_3cycle} shows some $L(k)$ curves, for the value of $\lambda$ used in Figure~\ref{fig:lyap_3cycle}. Here we used $N=500000$, to see whether noise-induced chaos can be observed. At the scales displayed here, we do observe the formation of the envelope. Therefore we conclude, that at these scales, the system behaves chaotically. From the envelope's slope, we estimate the value of the exponent to be about 0.4. However, at smaller spatial scales, the dynamics are dominated by the stochastic effects, and the signature of chaos (the envelope) cannot be detected. If we reduce the value of $N$, the stochastic effects become stronger and the envelope cannot be seen at any spatial scale. Results from the two methods, depicted in Figures~\ref{fig:lyap_3cycle} and \ref{fig:n_i_3cycle} both indicate chaotic behavior for this parameter choice, although they are not quantitatively consistent, with the first method returning a much lower Lyapunov exponent (obtained by reading off the graph for $N=500000$). Furthermore, the second method, which only employs knowledge of the time series, suggests that one must consider the balance between the two sources of perturbation in the system: chaos and the stochasticity. If the noise is too strong, then the signature of chaos cannot be detected. Moreover, one must consider the notion of spatial scale when studying these systems. Clearly more work is required to elucidate these features, and we are currently actively exploring some aspects of them.

The results found here are qualitatively similar to those found in previous studies on additive noise processes. For our multiplicative noise process, however, we noticed differences in the results obtained for the time-dependent exponent curves $L(k)$. The range of parameters ($N$,$\lambda$) for which noise-induced chaos was observed was smaller than found for the additive noise process. One possible explanation for this is that the heterogeneity of the stochasticity of a multiplicative noise process makes it more difficult to detect the signature of chaos. This is an interesting avenue for further work.
\begin{figure}
\begin{center}
\includegraphics[scale=0.9]{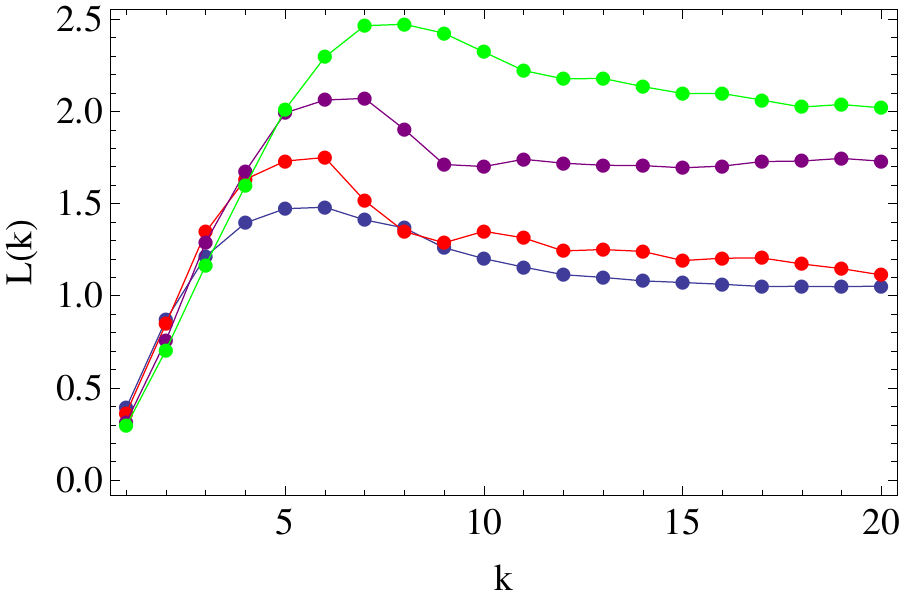}
\end{center}
\caption{Evolution of $L(k)$ for four shells $(2^{-(i+1)/2},2^{-i/2})$, $i=4,5,6,7$, from bottom to top, where the dots mark the values obtained for each integer $k$. Parameters are N=500000, $\lambda=3.8286$.}
\label{fig:n_i_3cycle}
\end{figure}
\subsection{The Ricker Map}
\label{sec:exp_map}
To demonstrate the general applicability of the method, we now turn to a second case of a one-dimensional map, which has the following form
\begin{equation}
x_{n+1}=x_n \textrm{exp} [ (r(1-x_n) ].
\end{equation}
In the literature, this map is sometimes referred to as the Ricker map \cite{ricker_54}, and has been much studied in the ecological literature (see e.g. Refs.~\cite{may_74,may_76,moll_08}).
In contrast to the logistic map, this map is not restricted to the unit interval. Indeed the maximum of the function, say $m_r$, occurs at $x=1/r$ and is found to be $(1/r)\textrm{exp}[r-1]$. However, for a particular choice of $r$ we can restrict the map to the unit interval by making a change of variables, $y_n=x_n/m_r$, such that the characteristics of the map (stability of the fixed point, bifurcation points, etc.) are preserved. The equation for the map for $y_n$ is
\begin{equation}
y_{n+1}=g(y_n)=y_n\textrm{exp} [ (r(1-m_ry_n) ].
\label{eq:exp_map}
\end{equation}
In Figure~\ref{fig:exp_map} we show both the unscaled and scaled function. Both maps have a stable fixed point for $0<r<2$.

The Markov chain process, and its mesoscopic description, can now be formulated as before, using Eq.~\eqref{binomial}, with a new choice of the function $p$. Figure~\ref{fig:moran_pdf} compares the stationary distribution of the Markov chain with simulation results from the stochastic difference equation for $r=1.95$. In Figure~\ref{fig:moran_pdf2} we show a similar plot, this time for $r=2.05$, for which the deterministic map has a stable 2-cycle.
\begin{figure}
\begin{center}
\includegraphics[scale=0.9]{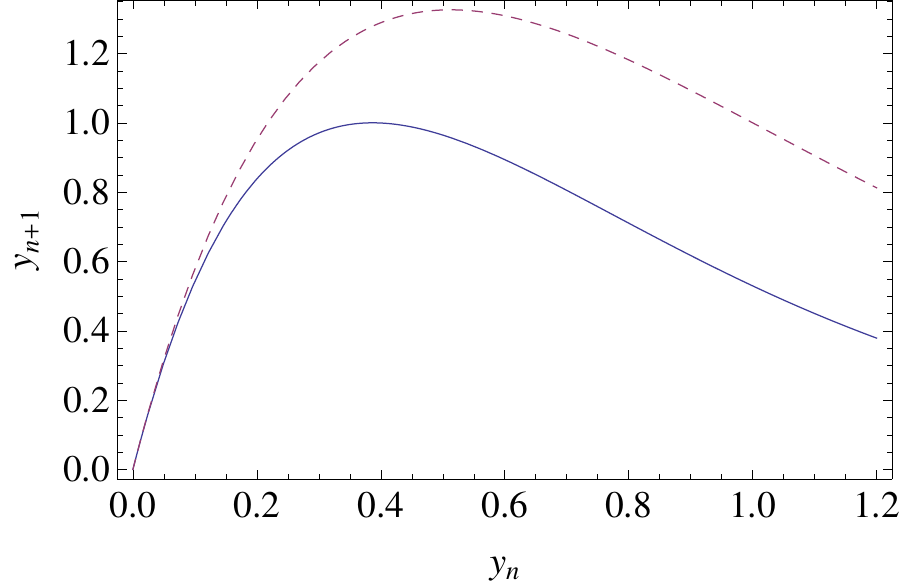}
\end{center}
\caption{Plot of the map given in Eq.~\eqref{eq:exp_map}. The dashed line is the original map, the solid line shows the rescaled map where the dynamics remain in the unit interval.}
\label{fig:exp_map}
\end{figure}
\begin{figure}
\begin{center}
\includegraphics[scale=0.9]{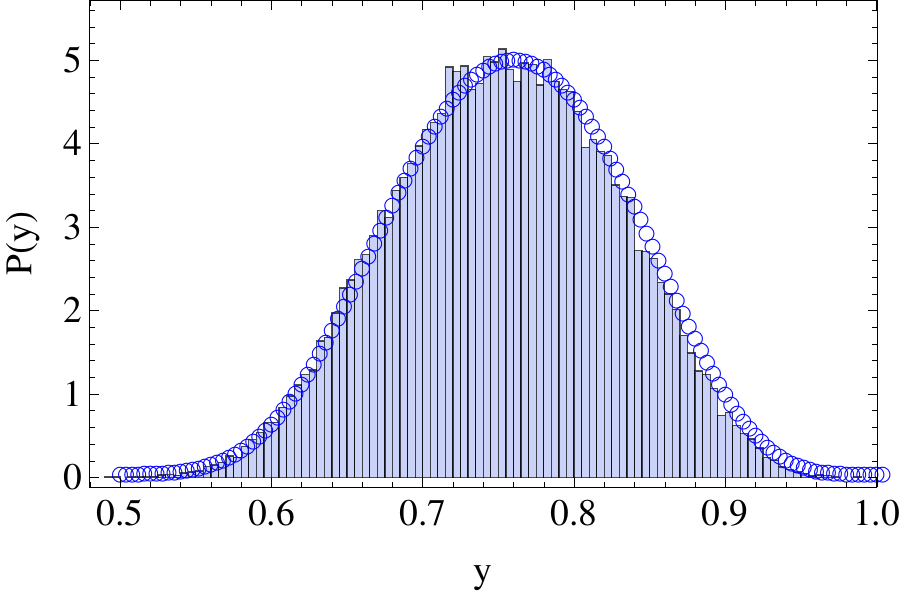}
\end{center}
\caption{Fluctuations around the fixed point for the Ricker map for $y$. Parameters chosen were $N=250$ and $r=1.95$. Results from the stochastic difference equation (bars) are compared with the stationary distribution of the Markov chain (circles). }
\label{fig:moran_pdf}
\end{figure}
\begin{figure}
\begin{center}
\includegraphics[scale=0.9]{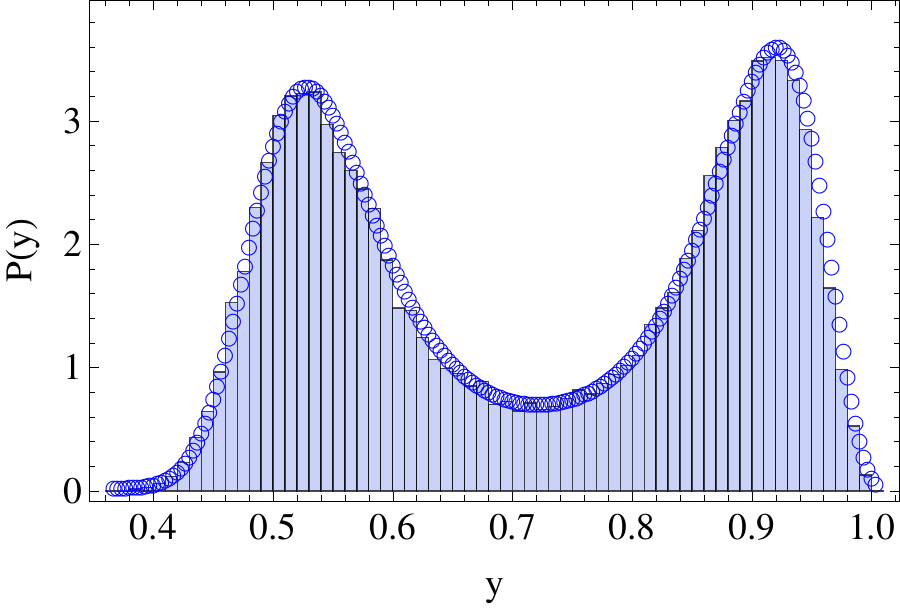}
\end{center}
\caption{Fluctuations around the 2-cycle for the Ricker map for $y$. Parameters chosen were $N=300$ and $r=2.05$. Results from the stochastic difference equation (bars) are compared with the stationary distribution of the Markov chain (circles).}
\label{fig:moran_pdf2}
\end{figure}
\section{Discussion}
\label{disc}
In this paper we have investigated the large $N$ limit of a Markov process defined by an $(N+1) \times (N+1)$ transition matrix $Q$. Over the last few years there have been many instances of carrying out such an investigation for continuous time Markov processes \cite{black_12,goutsias_13}. In this case one starts out from a master equation with $N$ discrete states, and using a diffusion-type approximation derives a Fokker-Planck equation. It is then possible to study the deterministic ($N \to \infty$) limit and the fluctuations about it, either within a linear noise approximation or within some other type of approximation which retains the multiplicative aspect of the noise \cite{mckane_13}.

However we are not aware of any previous studies that have carried out this program in the case where time is discrete. Of course, the application of the diffusion approximation to the Wright-Fisher model is well-known --- and was the original problem to which this approximation was applied \cite{fisher_22} --- but it is usual to make the time continuous as part of the diffusion approximation itself \cite{ewens_04}. Here we have not done this: we have made the state space continuous, but retain the discreteness of the time. As a consequence the equation corresponding to the Fokker-Planck equation is a differential-difference equation and what was a stochastic differential equation in the continuous-time case is now a stochastic difference equation.

Although our main motivation has been to set out the formalism in the discrete-time setting which parallels the well-known one when time is continuous, we chose the structure of the transition matrix $Q$ based on a biological application. Namely, we assumed that we were describing a system which at time $t$ consisted of $n$ identical individuals, and $(N-n)$ ``nulls'', that is, a system with a carrying capacity of $N$, which was only $n/N$ ``full''. This is equivalent to a Wright-Fisher model in genetics in which there are $n$ individual of type $A$ (i.e. having allele $A$) and $(N-n)$ individuals of type $B$ (i.e. having allele $B$) \cite{ewens_04}. The community at time $t+1$ is then created by sampling the community at time $t$ $N$ times. The main difference from the Wright-Fisher model was that we implemented competition for resources by making it less likely that we picked an individual, and more likely that we picked a null, as the number of individuals in the population grew. The interpretation in terms of sampling the $t$th generation to create the $(t+1)$th generation gave the $Q$ its binomial structure, but there may be other possible interpretations which would give a different form for $Q$. Hopefully these will naturally occur in particular applications; the specific form we assume here has the advantage that it gives us a definite structure to work with and so to develop the formalism and understand the results it leads to.

When studying the large $N$ behavior of such models, be they in discrete or continuous time, it is natural to first look at the system obtained when $N \to \infty$. This ``thermodynamic limit'' will ensure that all stochastic aspects of the model is lost and, not surprisingly, the outcome when time is discrete is a discrete map, the precise form of which depends on the choice of the matrix $Q$. We should stress (as we have done before in the continuous time case \cite{black_12,mckane_13}) that the map obtained by this procedure simply reflects the dynamics of the mean of the stochastic process; potentially there are an infinity of processes corresponding to this map, which differ in their higher moments. Thus simply ``adding noise'' to the map to get a stochastic version of the map --- as was done extensively in the past (see e.g. Refs~\cite{mayer_81,crutchfield_82,gao_99}) --- will give one of these infinity of models, and one with no particular microscopic motivation.

Starting from the model we have described above we obtained the logistic map in the $N \to \infty$ limit. This is not too surprising in itself; in the continuous time case the logistic differential equation is obtained. However, although the logistic differential equation can be solved exactly, and has a single attractor in the form of the fixed point, the logistic map famously has an extremely rich set of attractors. This crucial difference between the continuous and discrete time formulations allows us to explore how demographic noise and chaos interact in a controlled way, as we discuss again below. The simplest stochastic correction to the deterministic map is to include linear deviations from the attractors due to the demographic noise. In section 3 we obtained a general expression for the variance of the Gaussian pdf about a $n$-cycle, and showed this was in excellent agreement with simulations or the numerical solution of the stochastic difference equation. It is important to notice that although the noise is additive within this approximation, the procedure is quite different to simply adding noise to the deterministic map, since the noise correlations are determined by the underlying microscopic model. The general results for the variance about an $n$-cycle illustrate this: they depend on the parameter of the logistic map, $\lambda$, through the Jacobian $J$ and the noise strength $B$ evaluated on the $n$-cycle. 

We also studied the system in cases where there were significant deviations from the linearization described above. Since the linear approximation leads to a Gaussian pdf, this is equivalent to looking at situations in which the stationary pdf cannot be well-approximated by a Gaussian. We checked the accuracy of the stochastic difference equation against results found directly from the Markov chain. This is only practical for not too large values of $N$, because the computational cost of starting from the Markov chain becomes prohibitive quite quickly as $N$ increases. However, the excellent agreement we obtained for a range of $N$ gives us confidence in the predictions made from the stochastic difference equation for larger values of $N$. Since $N$ is simply a parameter in this equation, there is no extra cost in performing numerical calculations for arbitrarily large values of $N$.

One of the most interesting aspects of the discrete-time formalism, as compared with the more familiar case where time is continuous, is the nature of the Fokker-Planck equation. The usual truncation obtained by neglecting $\mathcal{O}(N^{-2})$ in the Kramers-Moyal expansion does not simply give two terms, but an infinite number. Essentially this is telling us that we are not using the correct variables to describe the stochastic process: in continuous-time processes the jumps $z(t) \to z(t+dt)$ are small, but in the discrete-time processes we have been investigating here the jumps $z_t \to z_{t+1}$ need not be small --- a fact already apparent from the corresponding deterministic map. What is small is the jump $p_t \to z_{t+1}$, which suggests that we should be working in part with the pdfs in the $p$ variable. If we do this in an appropriate way, then we can find a Fokker-Planck-like equation of a more familiar form. This formalism does not to the best of our knowledge appear in the literature, and deserves further study in its own right.

Since discrete maps with only a single degree of freedom can show chaotic behavior, we may use the formalism that we have developed to systematically investigate the nature of the chaotic state in stochastic systems which give rise to these maps in the deterministic limit. In Section 4, we estimated the Lyapunov exponent using two different methods which have been suggested in the literature. Past work was restricted to the case where noise was simply added to the deterministic map, and further work is required to understand how having intrinsic noise differs from this situation. Further work is also needed to assess the different ways in which Lyapunov exponents can be generalized to stochastic systems, and whether these give rise to significantly different conclusions for how stochasticity and chaos interact. It could also be interesting to compare the stationary probability distribution of the process (for various values of $N$) with the invariant measure of the corresponding deterministic map, to further study the effects of stochasticity in these systems. Numerical simulations show that, when the deterministic map is chaotic, the stationary probability distribution at finite $N$ is similar but smoother in appearance. In the future, we hope that more quantitative results could be obtained using our approach. More generally, many aspects of intrinsic noise in discrete-time systems remain to be investigated, from technical aspects relating to the possible ways of constructing the Markov chain, to the construction of practical models in ecology to generalizations to systems with many components. We hope to carry out and report on these questions in the future.

\medskip

\noindent \textbf{Acknowledgements} AJM wishes to thank the mathematical biology group at the University of Oxford for hospitality during the period when a significant portion of this work was carried out. We also thank Amos Maritan for useful correspondence. JDC and DF acknowledge support from Programs Prin2009 and Prin2012 financed by the Italian MIUR.

\medskip

\appendix

\section*{Appendix A: Derivation of the Kramers-Moyal expansion}
\label{App_1}
In this appendix we discuss the derivation of the Kramers-Moyal expansion for discrete-time stochastic processes of the kind that we are interested in and which are described in Section 2 of the main text. 

The form that we ultimately use, given by Eq.~(\ref{KM_new}), is of a different type to that found in textbooks \cite{gardiner_09,risken_89}. However let us begin by briefly reviewing the derivation of the conventional form as given by Eq.~(\ref{KM_conventional}). To derive this equation, we start from the Chapman-Kolmogorov equation which defines a Markov process. We shall suppress the dependence on the initial conditions $z_0$ at $t_0$, since they play no part in the derivation. Letting $z' = z - \Delta z$, the integrand of the integral appearing in Eq.~(\ref{CK_eqn}) may be written as
\[
P( \left[ z-\Delta z \right] + \Delta z,t+1 |z-\Delta z,t)P(z-\Delta z,t).
\]
Taylor expanding this, the integrand equals
\[
\sum^{\infty}_{\ell=0} \left( -1 \right)^{\ell} 
\frac{\left( \Delta z \right)^{\ell}}{\ell !}\,\frac{\partial^{\ell} }
{\partial z^{\ell}} \left\{ P(z + \Delta z,t+1 |z,t)P(z,t) \right\}.
\]
In the Chapman-Kolmogorov equation we had to integrate over $z'$. For fixed $z$, this is equivalent to integrating over $\Delta z$. We integrate $\Delta z$ from minus infinity to plus infinity, even though it is ``small''. This assumes that the integrand is very highly peaked and essentially gives no contribution to the integral apart from at very small $\Delta z$.

Introducing the jump moments defined by
\[
M_{\ell}(z) \equiv \int d(\Delta z) (\Delta z)^{\ell} P(z+\Delta z,t+1|z,t),
\]
we find that the Chapman-Kolmogorov equation reduces to Eq.~(\ref{KM_conventional}). Another form for the $M_{\ell}(z)$ is
\[
M_{\ell}(z)=\int dw\,(w-z)^{\ell} P(w,t+1|z,t),
\]
from which the form (\ref{defn_M}) given in the main text can be found.

However, as described in the main text, the Kramers-Moyal expansion (\ref{KM_conventional}) is not so useful. Essentially this stems from the fact that $\Delta z$ is not ``small'', and so the expansion cannot usefully be truncated after a finite number (that is, two) terms. We showed, in Eqs.~(\ref{FT_step1}) and (\ref{FT_step2}), that the first term in the expansion could be considerably simplified by taking the Fourier transform. We can also show this for the next term (the diffusion-like term). Its Fourier transform is
\begin{eqnarray*}
& & \frac{1}{2N}\,\sum_{\ell=0}^{\infty} 
\frac{(-1)^{\ell}}{\ell!}\,\int^{\infty}_{-\infty} dz\,e^{ikz}\,
\frac{\partial^{\ell + 2}}{\partial z^{\ell + 2}} \left[ (p-z)^{\ell}\,p(1-p)\,
P_t(z)\right] \nonumber \\
& & = \frac{1}{2N}\,\sum_{\ell=0}^{\infty} \frac{1}{\ell!}\,
\int^{\infty}_{-\infty} dz\,e^{ikz}\,
\left( ik \right)^{\ell + 2} \left(p - z \right)^{\ell}\,p(1-p)\,P_t(z) 
\nonumber \\
& & = \frac{1}{2N}\,\int^{\infty}_{-\infty} dz\,e^{ikz}\,e^{ik(p-z)}\,
\left( ik \right)^{2}\,p(1-p)\,P_t(z) \nonumber \\
& & = \frac{1}{2N}\,\int^{\infty}_{-\infty} dz\,e^{ikp}\,
\left( ik \right)^{2}\,p(1-p)\,P_t(z) \nonumber \\
& & = \frac{1}{2N}\,\int^{\infty}_{-\infty} dp\,e^{ikp}\,
\left( ik \right)^{2}\,p(1-p)\,\mathcal{P}_t(p) \nonumber \\
& & = \frac{1}{2N}\,\int^{\infty}_{-\infty} dp\,e^{ikp}\,
\frac{\partial^2}{\partial p^2}\left[ p(1-p) \mathcal{P}_t(p) \right].
\end{eqnarray*}
Taking the inverse Fourier transform of this expression gives the diffusion-like term in Eq.~(\ref{second_order_form}).

In order to be relatively concise, there are several aspects of the above derivation that we have treated in a rather informal manner. In the first three lines of the derivation $p$ is shorthand for $p(z)$. On the fourth line the integral takes the form $\int^{\infty}_{-\infty} f(p)\,P_{t}(z)\,dz$, where again $p=p(z)$, and $f(p)$ in this case is $p(1-p) e^{ikp}$. The change of variable used to obtain the next line may not be $1-1$ and one may have to split up the range of integration, and carry out different transformations in different ranges. For example, in the case of the logistic map, the range is split up so that for $0 \leq z \leq 1/2$, the transformation is $z=z_{-} \equiv [1 - \sqrt{1 - (4p/\lambda)]}/2$, whereas for $1/2 \leq z \leq 1$ it is $z=z_{+} \equiv [1 + \sqrt{1 - (4p/\lambda)]}/2$. The key point is that this has no effect on $f(p)$, since $p=\lambda z(1-z)$ takes on the same value whether $z=z_{-}$ or $z=z_{+}$. Thus the only relevant part of the transformation is $P_t(z) 
 dz = \mathcal{P}_t(p) dp$, where
\begin{displaymath}
\mathcal{P}_t(p) = \frac{1}{\lambda}\,
\left[ 1 - \frac{4p}{\lambda} \right]^{-1/2} 
\left\{ P_t(z=z_{-}) + P_t(z=z_{+}) \right\}.
\end{displaymath}

Having checked that the simplification occurs for $r=2$ (the diffusion-like term), we now show that it occurs to all orders in $r$. To do, this let us eliminate the jump moments $M_{\ell}$ in favor of the jump moments $J_r$ by substituting Eq.~(\ref{M_to_J}) into Eq.~(\ref{KM_conventional}) to find
\begin{eqnarray}
P_{t+1}(z) &=& \sum_{\ell=0}^{\infty} \frac{(-1)^{\ell}}{\ell!} 
\frac{\partial^{\ell}}{\partial z^{\ell}} \left[ \sum^{\ell}_{r=0}\, 
{\ell \choose r} \left( p - z \right)^{\ell - r} J_r(p) P_t(z)\right] 
\nonumber \\
&=& \sum_{\ell=0}^{\infty} \sum^{\ell}_{r=0}\,\frac{(-1)^{\ell}}{(\ell - r)!\,r!} 
\frac{\partial^{\ell}}{\partial z^{\ell}} \left[ \left( p - z \right)^{\ell - r} 
J_r(p) P_t(z)\right].
\label{KM_inter}
\end{eqnarray}
But now for any series $\sum_{\ell=0}^{\infty} \sum^{\ell}_{r=0}\,a_{r \ell} = \sum^{\infty}_{r=0}\,\sum_{\ell= r}^{\infty} a_{r \ell}$, and so
\begin{eqnarray}
P_{t+1}(z) &=& \sum_{r=0}^{\infty} \sum^{\infty}_{\ell=r}\,
\frac{(-1)^{\ell}}{(\ell - r)!\,r!} \frac{\partial^{\ell}}{\partial z^{\ell}} 
\left[ \left( p - z \right)^{\ell - r} J_r(p) P_t(z)\right] \nonumber \\
&=& \sum_{r=0}^{\infty}\,\frac{(-1)^{r}}{r!}\,\sum^{\infty}_{m=0}\,
\frac{(-1)^{m}}{m!} \frac{\partial^{r+m}}{\partial z^{r+m}} 
\left[ \left( p - z \right)^{m} J_r(p) P_t(z)\right].
\label{KM_inter_2}
\end{eqnarray}

We now take the Fourier transform of Eq.~(\ref{KM_inter_2}) to find:
\begin{eqnarray}
\tilde{P}_{t+1}(k) &=& \int^{\infty}_{-\infty} dz\,e^{ikz} P_{t+1}(z) \nonumber \\
&=& \sum_{r=0}^{\infty}\,\frac{(-1)^{r}}{r!}\,\int^{\infty}_{-\infty} dz\,e^{ikz}\,
\sum^{\infty}_{m=0}\,\frac{(-1)^{m}}{m!} \frac{\partial^{r+m}}{\partial z^{r+m}} 
\left[ \left( p - z \right)^{m} J_r(p) P_t(z)\right] \nonumber \\
&=& \sum_{r=0}^{\infty}\,\frac{1}{r!}\,\int^{\infty}_{-\infty} dz\,e^{ikz}\,
\sum^{\infty}_{m=0}\,\frac{1}{m!} \left( ik \right)^{r+m}\, 
\left[ \left( p - z \right)^{m} J_r(p) P_t(z)\right] \nonumber \\
&=& \sum_{r=0}^{\infty}\,\frac{1}{r!}\,\int^{\infty}_{-\infty} dz\,e^{ikz}\,
e^{ik(p-z)}\,\left( ik \right)^{r}\,J_r(p) P_t(z) \nonumber \\
&=& \sum_{r=0}^{\infty}\,\frac{1}{r!}\,\int^{\infty}_{-\infty} dz\,e^{ikp}\,
\left( ik \right)^{r}\,J_r(p) P_t(z) \nonumber \\
&=& \sum_{r=0}^{\infty}\,\frac{1}{r!}\,\int^{\infty}_{-\infty} dp\,e^{ikp}\,
\left( ik \right)^{r}\,J_r(p) \mathcal{P}_t(p) \nonumber \\
&=& \sum_{r=0}^{\infty}\,\frac{(-1)^r}{r!}\,\int^{\infty}_{-\infty} dp\,e^{ikp}\,
\frac{\partial^{r}}{\partial p^{r}}\,\left[ J_r(p) \mathcal{P}_t(p) \right]. 
\end{eqnarray}
Taking the inverse Fourier transform of this equation we find Eq.~(\ref{KM_new}).

\section*{Appendix B: The form of the jump moments $J_r$ for large $N$.}
\label{App_2}
The truncation of the Kramers-Moyal expansion (\ref{KM_new}) at the $r=2$ term, relies on showing that the $J_r$ fall off like $N^{-2}$ for $r > 2$. In Section 2, we have already shown that $J_1 = 0$ (and that $J_0 = 1$). For orientation, let us determine $J_2$:
\begin{eqnarray}
\left\langle \left[ z_{t+1} - p_t \right]^2 \right\rangle_{z_{t}=z} &=& 
N^{-2}\,\left\langle n^2_{t+1} \right\rangle_{n_{t}=m} - 
2N^{-1}p\,\left\langle n_{t+1} \right\rangle_{n_{t}=m} + p^2 \nonumber \\
&=& N^{-2}\,\sum_{n} n^2 P(n,t+1|m,t) - 2N^{-1}p\,\sum_{n} n P(n,t+1|m,t)
+ p^2 \nonumber \\
&=& N^{-2}\,\sum_{n} n^2 Q_{n m} - 2N^{-1}p\,\sum_{n} n Q_{n m} + p^2 \nonumber \\
&=& N^{-2}\left[ Np(1-p) + N^{2}p^{2} \right] - 2N^{-1}p\,Np + p^{2} \nonumber \\
&=& N^{-1}p(1-p). 
\label{J_2}
\end{eqnarray}

More generally in this appendix we want to investigate the form of $J_r$ for large $N$ and any value of $r$. Now,
\begin{eqnarray}
J_{r}(p) &=& \left\langle \left[ z_{t+1}-p_t \right]^r \right\rangle_{z_{t}=z} 
\nonumber \\
&=& \frac{1}{N^{r}}\,\left\langle 
\left[ n_{t+1} - (Np_t) \right]^r \right\rangle_{n_{t}=m} \nonumber \\
&=& \frac{1}{N^{r}}\,\sum^{r}_{s=0}\,{r \choose s}\,
\left( - 1 \right)^{s}\,\left( Np \right)^{r - s}\,\left\{ \sum_n n^s
Q_{n m} \right\}.
\label{Jp_inter}
\end{eqnarray}
If we denote the $s$th moment of the binomial distribution by $\mu_{s}(p)$,
then Eq.~(\ref{Jp_inter}) becomes
\begin{equation}
J_{r}(p) = \frac{1}{N^{r}}\,\sum^{r}_{s=0}\,{r \choose s}\,
\left( - 1 \right)^{s}\,\left( Np \right)^{r - s}\,\mu_{s}(p).
\label{Jp}
\end{equation}

So to calculate the $J_{r}(p)$, the moments of the binomial distribution need to be known. We have seen that $J_1(p)=0$ and $J_{2}(p)=N^{-1}p(1-p)$. We can find $J_{3}(p)$ from
\begin{eqnarray}
J_{3}(p) &=& \frac{1}{N^{3}}\,\left[ (Np)^3 - 3 (Np)^2 \mu_1 + 3 (Np) \mu_2
- \mu_3 \right] \nonumber \\
&=& - \frac{p(1-p)(1-2p)}{N^2},
\label{Jp_3}
\end{eqnarray}
which goes like $N^{-2}$ and so does not contribute to the generalization of the Fokker-Planck equation. We now show that all $J_{r}(p)$ for $r \geq 3$ fall off this fast.
We use properties of the factorial moments to prove this result. Recall that the factorial moments are defined by 
\begin{equation}
\nu_r(p) \equiv \langle n(n-1)\ldots(n-r+1) \rangle = \sum^{N}_{n=0}\,
n(n-1)\ldots(n-r+1)\,{N \choose n} p^n (1-p)^{N-n},
\label{factorial}
\end{equation}
and can be shown to be given by
\begin{equation}
\nu_r(p) = N(N-1)\ldots(N-r+1) p^r.
\label{fact_formula}
\end{equation}
The proof is straightforward. Define
\[
\phi(p,w) = \sum^{N}_{n=0}\,{N \choose n} p^n (1-p)^{N-n} w^n = 
\left[ pw + (1-p) \right]^N.
\]
Then differentiating $r$ times with respect to $w$ and setting $w=1$ gives $\nu_r(p)$, and so the result.

Now,
\[
\nu_r(p) =\langle n(n-1)\ldots(n-r+1) \rangle = \langle n \left[ n^{r-1} 
- A n^{r-2} + C_1 n^{r-3} + \ldots \right] \rangle,
\] 
where $A = 1 + \ldots + (r-1) = r(r-1)/2$ and where $C_1$ (like the constants $C_2$ and $C_3$ appearing below) is constant which need not concern us. Here we assume that $r \geq 3$. So we have,
\begin{eqnarray*}
\nu_r(p) &=& \langle n^r \rangle - A \langle n^{r-1} \rangle 
+ C_1 \langle n^{r-2} \rangle + \ldots \nonumber \\
\mathrm{and} \ \ \nu_{r-1}(p) &=& \langle n^{r-1} \rangle 
- C_2 \langle n^{r-2} \rangle + \ldots.
\end{eqnarray*}
These results imply that
\[
\langle n^r \rangle = \nu_r + A \langle n^{r-1} \rangle + \ldots =
\nu_r + A \nu_{r-1} + C_3 \nu_{r-2} + \ldots.
\]
We can now use the result (\ref{fact_formula}): $\nu_r(p) = (Np)^r -Ap (Np)^{r-1} + \mathcal{O}(N)^{r-2}$ to see that
\begin{equation}
\mu_r(p) \equiv \langle n^r \rangle = (Np)^r + A(1-p) (Np)^{r-1} + 
\mathcal{O}(N)^{r-2}.
\label{mom_result}
\end{equation}
for $r \geq 3$. Since $A=r(r-1)/2$ and $\mu_2(p) = (Np)^2 + Np(1-p)$, the result is also true for $r=0,1,2$ with the correction term being exactly zero.

Substituting Eq.~(\ref{mom_result}) into Eq.~(\ref{Jp}) gives
\begin{eqnarray}
J_{r}(p) &=& \frac{1}{N^{r}}\,\sum^{r}_{s=0}\,{r \choose s}\,
\left( - 1 \right)^{s}\,\left[ \left( Np \right)^{r} + A(1-p)
\left( Np \right)^{r - 1} + \mathcal{O}\left( N ^{r - 2} \right)\right]
\nonumber \\
&=& p^{r}\,\sum^{r}_{s=0}\,{r \choose s}\,\left( - 1 \right)^{s}\,
\left[ 1 + \frac{1}{2} s(s-1) (1-p) \left( Np \right)^{-1} + 
\mathcal{O}\left( N^{-2} \right) \right].
\label{Jp_large_N}
\end{eqnarray}
But 
\[
\sum^r_{s=0}\,{r \choose s}\,\left( - 1 \right)^{s} = 0, \ \ 
\mathrm{for\ } r \neq 0,
\]
and if $r > 2$
\[
\sum^r_{s=0}\,{r \choose s}\,\left( - 1 \right)^{s}\,s(s-1) = 0.
\]
Therefore Eq.~(\ref{Jp_large_N}) proves that $J_r(p)=\mathcal{O}(N^{-2})$ for 
$r > 2$.

\bibliographystyle{spmpsci}

\begin{thebibliography}{10}
\providecommand{\url}[1]{{#1}}
\providecommand{\urlprefix}{URL }
\expandafter\ifx\csname urlstyle\endcsname\relax
  \providecommand{\doi}[1]{DOI~\discretionary{}{}{}#1}\else
  \providecommand{\doi}{DOI~\discretionary{}{}{}\begingroup
  \urlstyle{rm}\Url}\fi

\bibitem{biancalani_10}
Biancalani, T., Fanelli, D., {Di Patti}, F.: Stochastic {Turing} patterns in
  the {Brusselator} model.
\newblock Phys. Rev. E \textbf{81}, 046215 (2010)

\bibitem{black_12}
Black, A.J., McKane, A.J.: Stochastic formulation of ecological models and
  their applications.
\newblock Trends in Ecology \& Evolution \textbf{27}, 337--345 (2012)

\bibitem{butler_09}
Butler, T., Goldenfeld, N.: Robust ecological pattern formation induced by
  demographic noise.
\newblock Phys. Rev. E \textbf{80}, 030902(R) (2009)

\bibitem{cencini_13}
Cencini, M., Vulpiani, A.: Finite size {Lyapunov exponent: review on
  applications}.
\newblock J. Phys. A \textbf{46}, 254019 (2013)

\bibitem{challenger_13}
Challenger, J.D., Fanelli, D., McKane, A.J.: Intrinsic noise and
  {discrete-time} processes.
\newblock Phys. Rev. E \textbf{88}, 040102(R) (2013)

\bibitem{crutchfield_82}
Crutchfield, J.P., Farmer, J.D., Huberman, B.A.: Fluctuations and simple
  chaotic dynamics.
\newblock Physics Reports \textbf{92}, 45--82 (1982)

\bibitem{ewens_04}
Ewens, W.J.: Mathematical Population Genetics. I. Theoretical Introduction.
\newblock Springer-Verlag, New York (2004)

\bibitem{fisher_22}
Fisher, R.A.: On the dominance ratio.
\newblock Proc. Roy. Soc. Edin. \textbf{42}, 321--341 (1922)

\bibitem{gao_94b}
Gao, J., Zheng, Z.: Direct dynamical test for deterministic chaos.
\newblock Europhys. Lett. \textbf{25}, 485--490 (1994)

\bibitem{gao_94a}
Gao, J., Zheng, Z.: Direct dynamical test for deterministic chaos and optimal
  embedding of a chaotic time series.
\newblock Phys. Rev. E \textbf{49}, 3807--3814 (1994)

\bibitem{gao_99}
Gao, J.B., Hwang, S.K., Liu, J.M.: When can noise induce chaos?
\newblock Phys. Rev. Lett. \textbf{82}, 1132 (1999)

\bibitem{gardiner_09}
Gardiner, C.W.: Handbook of Stochastic Methods, fourth edn.
\newblock Springer-Verlag, Berlin (2009)

\bibitem{gillespie_76}
Gillespie, D.T.: A general method for numerically simulating the stochastic
  time evolution of coupled chemical reactions.
\newblock J. Comp. Phys. \textbf{22}, 403--434 (1976)

\bibitem{gillespie_92book}
Gillespie, D.T.: Markov processes: An Introduction for Physical Scientists.
\newblock Academic Press, San Diego (1992)

\bibitem{godfray_89}
Godfray, H.C.J., Hassell, M.P.: Discrete and continuous insect populations in
  tropical environments.
\newblock Journal of Animal Ecology \textbf{58}, 153--174 (1989)

\bibitem{goutsias_13}
Goutsias, J., Jenkinson, G.: Markovian dynamics on complex reaction networks.
\newblock Physics Reports \textbf{529}, 199--264 (2013)

\bibitem{hassell_75}
Hassell, M.P.: {Density-dependence} in {single-species} populations.
\newblock Journal of Animal Ecology \textbf{44}, 283--295 (1975)

\bibitem{hassell_76}
Hassell, M.P., Comins, H.N.: Discrete time models for {two-species}
  competition.
\newblock Theoretical Population Biology \textbf{9}, 202--221 (1976)

\bibitem{may_74}
May, R.M.: Biological populations with nonoverlapping generations: stable
  points, stable cycles and chaos.
\newblock Science \textbf{186}, 645--647 (1974)

\bibitem{may_75}
May, R.M.: Biological populations obeying difference equations: stable points,
  stable cycles, and chaos.
\newblock J. Theor. Biol. \textbf{51}, 511--524 (1975)

\bibitem{may_76}
May, R.M., Oster, G.F.: Bifurcations and dynamic complexity in simple
  ecological models.
\newblock The American Naturalist \textbf{110}, 573--599 (1976)

\bibitem{mayer_81}
{Mayer-Kress}, G., Haken, H.: The influence of noise on the logistic model.
\newblock J. Stat. Phys. \textbf{26}, 149--171 (1981)

\bibitem{mckane_13}
McKane, A.J., Biancalani, T., Rogers, T.: Stochastic pattern formation and
  spontaneous polarisation: the linear noise approximation and beyond.
\newblock Bull. Math. Biol.  (2013)

\bibitem{mckane_05}
McKane, A.J., Newman, T.J.: Predator-prey cycles from resonant amplification of
  demographic stochasticity.
\newblock Phys. Rev. Lett. \textbf{94}, 218102 (2005)

\bibitem{moll_08}
Moll, J.D., Brown, J.S.: Competition and coexistence with multiple
  {life-history} stages.
\newblock The American Naturalist \textbf{171}, 839--843 (2008)

\bibitem{neubert_92}
Neubert, M.G., Kot, M.: The subcritical collapse of predator populations in
  {discrete-time} {predator-prey models}.
\newblock Mathematical Biosciences \textbf{110}, 45--66 (1992)

\bibitem{ott_93}
Ott, E.: Chaos in Dynamical Systems.
\newblock Cambridge University Press, Cambridge (1993)

\bibitem{packard_80}
Packard, N.H., Crutchfield, J.P., Farmer, J.D., Shaw, R.S.: Geometry from a
  time series.
\newblock Phys. Rev. Lett. \textbf{45}, 712--716 (1980)

\bibitem{reichl_80}
Reichl, L.E.: A Modern Course in Statistical Physics, second edn.
\newblock John Wiley and Sons, New York (1998)

\bibitem{ricker_54}
Ricker, W.: Stock and recruitment.
\newblock Journal of the Fisheries Research Board of Canada \textbf{11},
  559--623 (1954)

\bibitem{risken_89}
Risken, H.: The Fokker-Planck Equation, second edn.
\newblock Springer-Verlag, Berlin (1989)

\bibitem{strogatz_94}
Strogatz, S.H.: Nonlinear Dynamics and Chaos.
\newblock Perseus Publishing, Cambridge, Mass. (1994)

\bibitem{takens_81}
Takens, F.: Detecting strange attractors in turbulence.
\newblock In: D.A. Rand, L.S. Young (eds.) Dynamical Systems and Turbulence,
  Lecture Notes in Mathematics, vol. 898, pp. 366--381. Springer-Verlag, Berlin
  (1981)

\bibitem{kampen_07}
{van Kampen}, N.G.: Stochastic Processes in Physics and Chemistry, third edn.
\newblock Elsevier, Amsterdam (2007)

\bibitem{zunino_12}
Zunino, L., Soriano, M.C., Rosso, O.A.: Distinguishing chaotic and stochastic
  dynamics from time series by using a multiscale symbolic approach.
\newblock Phys. Rev. E \textbf{86}, 046210 (2012)

\end{thebibliography}

\end{document}